\documentclass{tlp}

\usepackage{amsmath}
\usepackage{amssymb}
\usepackage{hhline}
\usepackage{verbatim}
\usepackage{url}
\usepackage{stmaryrd}

\usepackage{array}
\usepackage{fixtitle}

\newcolumntype{.}{D{.}{.}{-1}}

\newcommand*{\Sharing}{\textup{\textsf{Sharing}}}
\newcommand*{\ASub}{\textup{\textsf{ASub}}}

\newcommand*{\VI}{\mathit{VI}}
\newcommand*{\SG}{\mathit{SG}}
\newcommand*{\SH}{\mathit{SH}}
\newcommand*{\sh}{\mathit{sh}}
\newcommand*{\sfl}{\mathit{d}}
\newcommand*{\sgfl}{\mathit{d}}
\newcommand*{\CR}{\mathit{CR}}
\newcommand*{\CS}{\mathit{CS}}
\newcommand*{\gf}{\mathit{gf}}
\newcommand*{\SHm}{{\mathit{SH}^\rho}}
\newcommand*{\PSD}{\mathit{PSD}}
\newcommand*{\ScozzariShPSh}{\mathsf{Sh}^\mathsf{PSh}}

\newcommand{\sbin}{\mathop{\mathrm{sbin}}\nolimits}

\newcommand*{\Bool}{\mathit{Bool}}

\newcommand*{\false}{\mathrm{false}}
\newcommand*{\Models}[2]{[#1]_{#2}}
\newcommand{\reduce}{\mathop{\mathrm{reduce}}\nolimits}
\newcommand*{\reduced}{{\mathrm{red}}}

\newtheorem{defn}{Definition}

\newcommand{\summary}[1]{\textrm{\textbf{\textup{#1}}}}

\newcommand{\defrel}[1]{\mathrel{\buildrel \mathrm{def} \over {#1}}}
\newcommand{\defeq}{\defrel{=}}

\newcommand*{\fund}[3]{\mathord{#1}\colon#2\rightarrow#3}

\newcommand{\sset}[2]{{\renewcommand{\arraystretch}{1.2}
                      \left\{\,#1 \,\left|\,
                               \begin{array}{@{}l@{}}#2\end{array}
                      \right.   \,\right\}}}

\renewcommand{\emptyset}{\mathord{\varnothing}}

\newcommand*{\sseq}{\subseteq}
\newcommand*{\Nsseq}{\nsubseteq}

\newcommand*{\union}{\cup}
\newcommand*{\inters}{\cap}
\newcommand*{\setdiff}{\setminus}
\newcommand*{\bigunion}{\bigcup}
\newcommand*{\Min}{\inplus}

\newcommand{\glb}{\mathop{\mathrm{glb}}\nolimits}
\newcommand{\lub}{\mathop{\mathrm{lub}}\nolimits}

\newcommand{\st}{\mathrel{.}}
\newcommand{\itc}{\mathrel{:}}

\newcommand*{\piff}{\mathrel{\leftrightarrow}}

\newcommand*{\Terms}{\mathord{\mathit{Terms}}}
\newcommand*{\Vars}{\mathord{\mathit{Vars}}}
\newcommand*{\vars}{\mathop{\mathit{vars}}\nolimits}
\newcommand*{\mvars}{\mathop{\mathit{mvars}}\nolimits}

\newcommand*{\Subst}{\mathit{Subst}}
\newcommand*{\Bind}{\mathit{Bind}}

\newcommand*{\rel}{\mathop{\mathrm{rel}}\nolimits}
\newcommand*{\irel}{\mathop{\overline{\mathrm{rel}}}\nolimits}
\newcommand*{\bin}{\mathop{\mathrm{bin}}\nolimits}
\newcommand*{\aexists}{\mathop{\mathrm{aexists}}\nolimits}
\newcommand*{\mgu}{\mathop{\mathrm{mgu}}\nolimits}
\newcommand*{\amgu}{\mathop{\mathrm{amgu}}\nolimits}

\newcommand{\china}{\textmd{\textsc{China}}}

\newcommand*{\Pos}{\textit{Pos}}
\newcommand*{\Def}{\mathit{Def}}

\newcommand*{\Free}{\textit{Free}}
\newcommand*{\GF}{\textit{GF}}

\newcommand*{\Pattern}{\mathop{\mathrm{Pattern}}\nolimits}

\newcommand*{\HofN}{{\cH_{\mkern-3mu\cN}}}
\newcommand*{\CLPHofN}{\textup{CLP($\HofN$)}}

\newcommand*{\CLPR}{CLP($\cR$)}

\newcommand*{\cD}{\mathcal{D}}

\newcommand*{\cH}{\mathcal{H}}

\newcommand*{\cN}{\mathcal{N}}

\newcommand*{\cR}{\mathcal{R}}

\newcommand*{\Cplusplus}{{C\nolinebreak[4]\hspace{-.05em}\raisebox{.4ex}{\tiny\bf ++}}}

\newcommand*{\SFL}{\mathit{SFL}}
\newcommand*{\PSDFL}{\SFL_2}
\newcommand*{\SGFL}{\mathit{SGFL}}
\newcommand*{\PSDGFL}{\SGFL_2}
\newcommand*{\free}{\mathit{free}}
\newcommand*{\gfree}{\mathit{gfree}}
\newcommand*{\lin}{\mathit{lin}}
\newcommand*{\ind}{\mathit{ind}}

\begin{document}
\title[Enhanced Sharing Analysis Techniques: A Comprehensive Evaluation]
      {Enhanced Sharing Analysis Techniques:\\ A Comprehensive Evaluation}
\author[R. Bagnara, E. Zaffanella, and P. M. Hill]
       {ROBERTO BAGNARA, ENEA ZAFFANELLA\thanks{The work of the first
           and second authors has been partly supported by MURST projects
           ``Certificazione automatica di programmi
                      mediante interpretazione astratta''
           and
           ``Interpretazione astratta, sistemi di tipo
                      e analisi control-flow.''}
       \affiliation
       Department of Mathematics,
       University of Parma,
       Italy \\
       \email{ \{bagnara,zaffanella\}@cs.unipr.it}
       \and PATRICIA M. HILL\thanks{This work was partly supported
                                 by EPSRC under grant GR/M05645.}
       \affiliation
       School of Computing,
       University of Leeds,
       Leeds, U.K. \\
       \email{hill@comp.leeds.ac.uk}
}
\maketitle

\begin{abstract}
$\Sharing$, an abstract domain developed by D.~Jacobs and A.~Langen
for the analysis of logic programs, derives useful aliasing information.
It is well-known that a commonly used core of techniques, such as the
integration of $\Sharing$ with freeness and linearity information,
can significantly improve the precision of the analysis.
However, a number of other proposals for refined domain combinations
have been circulating for years.
One feature that is common to these proposals is that they do
not seem to have undergone a thorough experimental evaluation
even with respect to the expected precision gains.
In this paper we experimentally evaluate:
helping $\Sharing$ with the definitely ground variables found
using $\Pos$,
the domain of positive Boolean formulas;
the incorporation of explicit structural information;
a full implementation of the reduced product of $\Sharing$ and $\Pos$;
the issue of reordering the bindings in the computation of the abstract $\mgu$;
an original proposal for the addition of a new mode recording the
set of variables that are deemed to be ground or free;
a refined way of using linearity to improve the analysis;
the recovery of hidden information in the combination of $\Sharing$
with freeness information.
Finally, we discuss the issue of whether tracking compoundness
allows the computation of more sharing information.
\end{abstract}


\section{Introduction}

In the execution of a logic program, two variables are \emph{aliased}
or \emph{share} at some program point if they are bound to terms that
have a common variable.  Conversely, two variables are
\emph{independent} if they are bound to terms that have no variables
in common.
Thus by providing information about possible variable aliasing,
we also provide information about definite variable independence.
In logic programming,
a knowledge of the possible aliasing (and hence definite
independence) between variables has some important applications.

Information about variable aliasing is essential
for the efficient exploitation of
AND-parallelism~\cite{BuenodlBH94,BuenodlBH99,ChangDdG85,HermenegildoG90,%
HermenegildoR95,JacobsL92,MuthukumarH92}.
Informally, two atoms in a goal are executed in parallel if,
by a mixture of compile-time and run-time checks, it can
be guaranteed that they do not share any variable.
This implies the absence of \emph{binding conflicts} at run-time:
it will never happen that the processes associated to the
two atoms try to bind the same variable.

Another significant application is \emph{occurs-check reduction}
\cite{CrnogoracKS96,Sondergaard86}.
It is well-known that many implemented logic programming
languages (e.g., almost all Prolog systems) omit the
\emph{occurs-check} from the unification procedure.
Occurs-check reduction amounts to identifying  the unifications
where such an omission is safe, and, for this purpose, information
on the possible aliasing of program variables is crucial.

Aliasing information can also be used indirectly in
the computation of other interesting program properties.
For instance, the precision with which freeness information
can be computed depends,
in a critical way,
on the precision with which aliasing can be tracked
\cite{BruynoogheCM94,CodishDFB93,File94,KingS94,Langen90th,MuthukumarH91}.

In addition to these well-known applications, a recent line of research
has shown that aliasing information can be exploited
in \emph{Inductive Logic Programming} (ILP).
Several optimizations have been proposed for speeding up the refinement
of inductively defined predicates in ILP systems
\cite{BlockeelDJVVL00,SantosCostaSC00}.
It has been observed that the applicability of some of these
optimizations, formulated
in terms of syntactic conditions on the considered predicate,
could be recast as tests on variable aliasing
\cite[Appendix D]{BlockeelDJVVL00}.

$\Sharing$, a domain due to D.~Jacobs and A.~Langen
\cite{JacobsL89,JacobsL92,Langen90th}, is based on the concept
of \emph{set-sharing}.
An element of the $\Sharing$ domain, which is a set of \emph{sharing-groups}
(i.e., a set of sets of variables),
represents information on groundness,\footnote{
A variable is \emph{ground} if it is bound to a term containing no variables,
it is \emph{compound} if it is bound to a non-variable term,
it is \emph{free} if it is not compound,
it is \emph{linear} if it is bound to a term
that does not contain multiple occurrences of a variable.
}
groundness dependencies,
possible aliasing, and more complex \emph{sharing-dependencies}
among the variables that are involved in the execution
of a logic program \cite{BagnaraHZ97b,BagnaraHZ02TCS,BuenodlBH94,BuenodlBH99}.

Even though $\Sharing$ is quite precise,
it is well-known that more precision is attainable
by combining it with other domains.
Nowadays, nobody would seriously consider performing sharing analysis
without exploiting the combination of aliasing information
with groundness and linearity information.
As a consequence, expressions such as `sharing information',
`sharing domain' and `sharing analysis' usually capture
groundness, aliasing, linearity and quite often also freeness.
Notice that this idiom is nothing more than a historical accident:
as we will see in the sequel,
compoundness and other kinds of structural information
could also be included in the collective term `sharing information'.

As argued informally by H.~S{\o}ndergaard \cite{Sondergaard86},
linearity information can be suitably exploited to improve the accuracy
of a sharing analysis.
This observation has been formally applied in \cite{CodishDY91}
to the specification of the abstract $\mgu$ operator for $\ASub$,
a sharing domain based on the concept of \emph{pair-sharing}
(i.e., aliasing and linearity information is encoded by a set of pairs
of variables).
A similar integration with linearity for the domain $\Sharing$
was proposed by Langen in his PhD thesis \cite{Langen90th}.
The synergy attainable from the integration between aliasing and
freeness information was pointed out
by K.~Muthukumar and M.~Hermenegildo~\cite{MuthukumarH92}.
Building on these works, W.~Hans and S.~Winkler \cite{HansW92} proposed
a combined integration of freeness and linearity information with sharing,
but small variations (such as the one we will present as the starting point
for our work) have been developed
by M.~Bruynooghe et al.~\cite{BruynoogheC93,BruynoogheCM94}.

There have been a number of other proposals for more refined combinations
which have the potential for improving the precision of the
sharing analysis over and above that obtainable
using the classical combinations of $\Sharing$ with linearity and freeness.
These include
the implementation of more powerful abstract semantic operators
(since it is well-known that the commonly used ones
are sub-optimal) and/or the integration with other domains.
Not one of these proposals seem
to have undergone a thorough experimental evaluation,
even with respect to the expected precision gains.
The goal of this paper is to systematically study these enhancements
and provide a uniform theoretical presentation together with an
extensive experimental evaluation that will give a strong
indication of their impact on the accuracy of the sharing
information.

Our investigation is primarily from the point of view of precision.
Reasonable efficiency is also clearly of interest
but this has to be secondary to the question as to whether
precision is significantly improved: only if this is established,
should better implementations be researched.
One of the investigated enhancements is the integration of
explicit structural information in the sharing analysis
and an important contribution of this paper is that it shows
both the feasibility and the positive impact of this combination.

Note that, regardless of its practicality,
 any feasible sharing analysis technique that offers good precision may be
valuable.
While inefficiency may prevent its adoption in production analyzers,
it can help in assessing the precision of the more competitive techniques.

The present paper, which is an improved and extended version
of~\cite{BagnaraZH00}, is structured as follows.
In Section \ref{sec:preliminaries}, we define some notation
and recall the definitions of the domain $\Sharing$
and its standard integration with freeness and linearity
information denoted as $\SFL$.
In Section \ref{sec:experimental-evaluation}, we briefly describe
the $\china$ analyzer, the benchmark suite and the methodology
we follow in the experimental evaluations.
In each of the next seven sections,
we describe and experimentally evaluate different enhancements
and precision optimizations for the domain $\SFL$.
Section~\ref{sec:simplepos}
considers a simple combination of $\Pos$  with  $\SFL$;
Section~\ref{sec:struct}
investigates the effect of including explicit structural
information by means of the  $\Pattern(\cdot)$ construction;
Section~\ref{sec:binding-ordering}
discusses possible heuristics for reordering
the bindings so as to maximize the precision of $\SFL$;
Section~\ref{sec:enhancedpos}
studies the implementation of a more precise combination
between $\Pos$ and $\SFL$;
Section~\ref{sec:ground-or-free}
describes a new mode `ground or free' to be included in $\SFL$;
Section~\ref{sec:enhancedlin} and Section~\ref{sec:enhancedfree}
study the possibility of improving the exploitation of
the linearity and freeness information already encoded in $\SFL$.
In Section~\ref{sec:compoundness} we discuss (without an experimental
evaluation) whether compoundness information can be useful
for precision gains.
Section~\ref{sec:conclusion}
concludes with some final remarks.

\section{Preliminaries}
\label{sec:preliminaries}

For any set $S$, $\wp(S)$ denotes the powerset of $S$.
For ease of presentation, we assume there is a
finite set of variables of interest denoted by $\VI$.
If $t$ is a syntactic object then $\vars(t)$ and $\mvars(t)$
denote the set and the multiset of variables in $t$, respectively.
If $a$ occurs more than once in a multiset $M$
we write $a \Min M$.
We let $\Terms$ denote the set of first-order terms over $\VI$.
$\Bind$ denotes the set of equations of the form $x = t$
where $x \in \VI$ and $t \in \Terms$ is distinct from $x$.
Note that we do not impose the occurs-check condition
$x \notin\nobreak \vars(t)$, since we target the analysis of
Prolog and CLP systems possibly omitting this check.
The following simplification of the standard definitions
for the $\Sharing$ domain~\cite{CortesiF99,HillBZ98b,JacobsL92}
assumes that the set of variables of interest
is always given by $\VI$.\footnote{Note
that, during the analysis process, the set of variables of interest
may expand (when solving the body of a clause) and contract
(when abstract descriptions are projected onto the variables occurring
in the head of a clause).
However, at any given time the set of variables of interest is fixed.
By consistently denoting this set by $\VI$,
we simplify the presentation, since we can omit the set of variables
of interest to which an abstract description refers.}
\begin{defn} \summary{(The \emph{set-sharing} domain $\SH$.)}
The set $\SH$ is defined by
\begin{align*}
  \SH &\defeq \wp(\SG),
\intertext{%
where the set of \emph{sharing-groups} $\SG$ is given by
}
  \SG &\defeq \wp(\VI) \setdiff \{ \emptyset \}.
\end{align*}
\end{defn}
$\SH$ is ordered by subset inclusion.
Thus the $\lub$ and $\glb$ of the domain are
set union and intersection, respectively.

\begin{defn} \summary{(Abstract operations over $\SH$.)}
\label{def:abs-funcs}
The \emph{abstract existential quantification} on $\SH$
causes an element of $\SH$ to ``forget everything''
about a subset of the variables of interest.
It is encoded by the binary function
$\fund{\aexists}{\SH \times \wp(\VI)}{\SH}$ such that,
for each $\sh \in \SH$ and $V \in \wp(\VI)$,
\[
  \aexists(\sh, V)
    \defeq
      \bigl\{\,
        S \setdiff V
      \bigm|
        S \in \sh, S \setdiff V \neq \emptyset
      \,\bigr\}
        \union
          \bigl\{\, \{x\} \bigm| x \in V \,\bigr\}.
\]

For each $\sh \in \SH$ and each $V \in \wp(\VI)$,
the extraction of the
\emph{relevant component of $\sh$ with respect to $V$}
is given by the function
$\fund{\rel}{\wp(\VI)\times\SH}{\SH}$
defined as
\[
  \rel(V, \sh)
    \defeq
      \{\, S \in \sh \mid S \inters V \neq \emptyset \,\}.
\]

For each $\sh \in \SH$ and each $V \in \wp(\VI)$,
the function $\fund{\irel}{\wp(\VI)\times\SH}{\SH}$
gives the \emph{irrelevant component of $\sh$ with respect to $V$}.
It is defined as
\[
  \irel(V, \sh)
    \defeq \sh \setdiff \rel(V,\sh).
\]

The function
$\fund{(\cdot)^\star}{\SH}{\SH}$,
also called \emph{star-union},
is given,
for each $\sh \in \SH$, by
\[
  \sh^\star
    \defeq
      \biggl\{\,
        S \in \SG
      \biggm|
        \exists n \geq 1
          \st
            \exists T_1, \ldots, T_n \in \sh
              \st S = \bigunion_{i=1}^n T_i
      \,\biggr\}.
\]

For each $\sh_1, \sh_2 \in \SH$,
the function
$\fund{\bin}{\SH\times\SH}{\SH}$,
called \emph{binary union},
is given by
\[
  \bin(\sh_1, \sh_2)
    \defeq
      \{\,
        S_1 \union S_2
      \mid
        S_1 \in \sh_1,
        S_2 \in \sh_2
      \,\}.
\]

We also use the \emph{self-bin-union} function
$\fund{\sbin}{\SH}{\SH}$,
which is given, for each $\sh \in \SH$, by
\[
  \sbin(\sh)
    \defeq \bin(\sh, \sh).
\]

The function $\fund{\amgu}{\SH\times\Bind}{\SH}$
captures the effect of a binding on an element of $\SH$.
Assume $(x = t) \in \Bind$, $\sh \in \SH$,
$V_x = \{x\}$, $V_t = \vars(t)$, and $V_{xt} = V_x \union V_t$.
Then
\begin{equation}
\label{eq:amgu-std}
  \amgu(\sh, x = t)
    \defeq
      \irel(V_{xt},\sh)
        \union \bin\bigl(\rel(V_x, \sh)^\star, \rel(V_t, \sh)^\star\bigr).
\end{equation}
\end{defn}

We now briefly recall the standard integration of set-sharing
with freeness and linearity information. These properties are
each represented by a set of variables, namely those variables
that are bound to terms that definitely enjoy the given property.
These sets are partially ordered by \emph{reverse} subset inclusion
so that the $\lub$ and $\glb$ operators are given by
set intersection and union, respectively.

\begin{defn} \summary{(The domain $\SFL$.)}
Let $F \defeq \wp(\VI)$ and $L \defeq \wp(\VI)$
be partially ordered by reverse subset inclusion.
The domain $\SFL$ is defined by the Cartesian product
\[
  \SFL \defeq \SH \times F \times L
\]
ordered by the component-wise extension of the orderings
defined on the three subdomains.
\end{defn}
A complete definition would explicitly deal with the set
of variables of interest $\VI$.
We could even define an equivalence relation on $\SFL$ identifying
the bottom element $\bot \defeq \langle \emptyset, \VI, \VI \rangle$
with all the elements corresponding to
an impossible concrete computation state:
for example, elements $\langle \sh, f, l \rangle \in \SFL$ such that
$f \Nsseq \vars(\sh)$ (because a free variable does share with itself)
or $\VI \setdiff \vars(\sh) \Nsseq l$ (because variables that cannot
share are also linear).
Note however that these and other similar spurious elements
rarely occur in practice and
cannot compromise the correctness of the results.

In a bottom-up abstract interpretation framework,
such as the one we focus on, abstract unification is
the only critical operation.
Besides unification,
the analysis depends on the `merge-over-all-paths' operator,
corresponding to the lub of the domain,
and the abstract projection operator, which can be defined
in terms of an abstract existential quantification operator.

\begin{defn} \summary{(Abstract operations over $\SFL$.)}
\label{def:abs-funcs-SFL}
The \emph{abstract existential quantification} on $\SFL$
is encoded by the binary function
$\fund{\aexists}{\SFL\times\wp(\VI)}{\SFL}$ such that,
for each $\sfl = \langle \sh, f, l \rangle \in \SFL$ and $V \in \wp(\VI)$,
\[
  \aexists(\sfl, V)
    \defeq
      \bigl\langle
        \aexists(\sh, V),
        f \union V,
        l \union V
      \bigr\rangle.
\]

For each $\sfl = \langle \sh, f, l \rangle \in \SFL$,
we define the following predicates.
The predicate $\fund{\ind_{\sfl}}{\Terms \times \Terms}{\Bool}$
expresses definite independence of terms.
Two terms $s,t \in \Terms$ are \emph{independent in $\sfl$}
if and only if $\ind_{\sfl}(s, t)$ holds,
where
\begin{align*}
  \ind_{\sfl}(s, t)
    &\defeq
      \Bigl(
        \rel\bigl( \vars(s), \sh \bigr)
          \inters
          \rel\bigl( \vars(t), \sh \bigr)
        = \emptyset
       \Bigr). \\
\intertext{%
A term $t \in \Terms$ is \emph{free in $\sfl$} if and only if the predicate
$\fund{\free_{\sfl}}{\Terms}{\Bool}$ holds for $t$, that is,
}
  \free_{\sfl}(t)
    &\defeq
      \bigl(
        \exists x \in \VI \st x = t \land x \in f
      \bigr). \\
\intertext{%
A term $t \in \Terms$ is \emph{linear in $\sfl$} if and only if
$\lin_{\sfl}(t)$,
where $\fund{\lin_{\sfl}}{\Terms}{\Bool}$ is given by
}
  \lin_{\sfl}(t)
    &\defeq
       \bigl( \vars(t) \sseq l \bigr) \\
      &\qquad \land \bigl(
               \forall x,y \in \vars(t)
                 \itc
                   x = y \lor \ind_{\sfl}(x, y)
             \bigr) \\
      &\qquad \land \Bigl(
               \forall x \in \vars(t)
                 \itc
                   x \Min \mvars(t) \Rightarrow x \notin \vars(\sh)
             \Bigr).
\end{align*}

The function $\fund{\amgu}{\SFL\times\Bind}{\SFL}$
captures the effects of a binding on an element of $\SFL$.
Let $(x = t) \in \Bind$ and $\sfl = \langle \sh, f, l \rangle \in \SFL$.
Let also $V_x = \{x\}$, $V_t = \vars(t)$, $V_{xt} = V_x \union V_t$,
$R_x = \rel(V_x, \sh)$ and $R_t = \rel(V_t, \sh)$.
Then
\begin{align*}
  \amgu\bigl(\sfl, x = t\bigr)
    &\defeq
      \langle \sh', f', l' \rangle,
\end{align*}
where
\begin{align*}
  \sh' &\defeq \irel(V_{xt},\sh) \union \bin\bigl(S_x, S_t\bigr);\\
  S_x  &\defeq \begin{cases}
             R_x,	&\text{if $\free_{\sfl}(x)
                                  \lor
                                  \free_{\sfl}(t)
                                  \lor
                                  \bigl(
                                    \lin_{\sfl}(t)
                                    \land
                                    \ind_{\sfl}(x, t)
                                  \bigr)$;} \\
             R_x^\star, &\text{otherwise;}
           \end{cases}\\
  S_t  &\defeq \begin{cases}
            R_t,       &\text{if $\free_{\sfl}(x)
                                 \lor \free_{\sfl}(t)
                                 \lor \bigl(
                                        \lin_{\sfl}(x)
                                        \land
                                        \ind_{\sfl}(x, t)
                                      \bigr)$;} \\
            R_t^\star, &\text{otherwise;}
          \end{cases}\\
  f'   &\defeq \begin{cases}
            f, &\text{if $\free_{\sfl}(x) \land \free_{\sfl}(t)$;} \\
            f \setdiff \vars(R_x), &\text{if $\free_{\sfl}(x)$;} \\
            f \setdiff \vars(R_t), &\text{if $\free_{\sfl}(t)$;} \\
            f \setdiff \vars(R_x \union R_t),
                                   &\text{otherwise;} \\
          \end{cases} \\
  l'   &\defeq \bigl( \VI \setdiff \vars(\sh') \bigr) \union f' \union l''; \\
  l''  &\defeq \begin{cases}
            l \setdiff \bigl(
                          \vars(R_x) \inters \vars(R_t)
                        \bigr),    &\text{if $\lin_{\sfl}(x)
                                              \land
                                              \lin_{\sfl}(t)$;} \\
            l \setdiff \vars(R_x), &\text{if $\lin_{\sfl}(x)$;} \\
            l \setdiff \vars(R_t), &\text{if $\lin_{\sfl}(t)$;} \\
            l \setdiff \vars(R_x \union R_t),
                                   &\text{otherwise.} \\
          \end{cases}
\end{align*}
\end{defn}

This specification of the abstract unification operator
is equivalent (modulo the lack of the explicit structural information
provided by \emph{abstract equation systems})
to that given in~\cite{BruynoogheCM94},
provided $x \notin \vars(t)$.
Indeed, as done in all the previous papers on the subject,
in~\cite{BruynoogheCM94} it is assumed that the analyzed language
does perform the occurs-check.
As a consequence, whenever considering a \emph{definitely cyclic binding},
that is a binding $x=t$ such that $x \in \vars(t)$,
the abstract operator can detect the definite failure
of the concrete computation and thus return the bottom element
of the domain.
Such an improvement would not be safe in our case,
since we also consider languages possibly omitting the occurs-check.
However, when dealing with definitely cyclic bindings,
the specification given by the previous definition
can still be refined as follows.

\begin{defn} \summary{(Improvement for definitely cyclic bindings.)}
\label{def:abs-funcs-SFL-cyclic}
Consider the specification of the abstract operations over $\SFL$
given in Definition~\ref{def:abs-funcs-SFL}.
Then, whenever $x \in \vars(t)$, the computation of the new
sharing component $\sh'$ can be
replaced by the following.\footnote{Note that, in this special case,
it also holds that $\free_{\sfl}(t)=\false$ and
$\ind_{\sfl}(x,t) = (R_x = \emptyset)$.}
\begin{align*}
  \sh' &\defeq \irel(V_{xt},\sh) \union \bin\bigl(S_x, \CS_t\bigr),\\
\intertext{where}
  \CS_t &\defeq \begin{cases}
             \CR_t,       &\text{if $\free_{\sfl}(x)$;} \\
             \CR_t^\star, &\text{otherwise;}
           \end{cases} \\
  \CR_t &\defeq \rel\bigl(\vars(t) \setdiff \{x\}, \sh\bigr).
\end{align*}
\end{defn}
This enhancement, already implemented in the \china{} analyzer,
is the rewording of a similar one proposed in~\cite{Bagnara97th}
for the domain $\Pos$ in the context of groundness analysis.
Its net effect is to recover some groundness and sharing dependencies
that are unnecessarily lost when using the standard operators.

The domain $\SH$ captures \emph{set-sharing}.
However, the property we wish to detect is \emph{pair-sharing}
and, for this, it has been shown in~\cite{BagnaraHZ02TCS}
that $\SH$ includes unwanted redundancy.
The same paper introduces an upper-closure operator $\rho$ on $\SH$
and the domain $\PSD \defeq \rho(\SH)$,
which is the weakest abstraction of $\SH$ that is
\emph{as precise as $\SH$}
as far as tracking groundness and pair-sharing is concerned.\footnote{%
The name $\PSD$, which stands for \emph{Pair-Sharing Dependencies},
was introduced in~\cite{ZaffanellaHB99}.
All previous papers, including \cite{BagnaraHZ02TCS},
denoted this domain by $\SHm$.}
A notable advantage of $\PSD$ is that we can replace the star-union
operation in the definition of the $\amgu$ by self-bin-union
without loss of precision. In particular,
in~\cite{BagnaraHZ02TCS} it is shown that
\begin{equation}
\label{eq:amgu-rho}
  \amgu(\sh, x = t)
    =_\rho
      \irel(V_{xt},\sh)
        \union \bin\Bigl(\sbin\bigl(\rel(V_x, \sh)\bigr),
                         \sbin\bigl(\rel(V_t, \sh)\bigr)
                   \Bigr),
\end{equation}
where the notation
$\sh_1 =_\rho \sh_2$
means
$\rho(\sh_1) = \rho(\sh_2)$.

It is important to observe that the complexity of the $\amgu$ operator
on $\SH$ (\ref{eq:amgu-std}) is exponential in the number
of sharing-groups of $\sh$.
In contrast, the operator on $\PSD$ (\ref{eq:amgu-rho})
is $O\bigl(|\sh|^4\bigr)$.
Moreover, checking whether a fixpoint has been reached by testing
$\sh_1 =_\rho \sh_2$ has complexity $O\bigl(|\sh_1|^3+|\sh_2|^3\bigr)$.
Practically speaking, very often this makes the difference between
thrashing and termination of the analysis in reasonable time.

The above observations on $\SH$ and $\PSD$ can be generalized
to apply to the domain combinations $\SFL$ and
$\PSDFL \defeq \PSD \times F \times L$.
In particular, $\PSDFL$ achieves the same precision as $\SFL$
for groundness, pair-sharing, freeness and linearity
and the complexity of the corresponding abstract unification operator
is polynomial.
For this reason, all the experimental work in this paper,
with the exception of part of the one described
in Section~\ref{sec:enhancedpos},
has been conducted using the $\PSDFL$ domain.

\section{Experimental Evaluation}
\label{sec:experimental-evaluation}

Since the main purpose of this paper is to provide
an experimental measure of the precision gains
that might be achieved by enhancing a standard sharing analysis
with several new techniques we found in the literature,
it is clear that the implementation of the various domain combinations
was a major part of the work.
However, so as to adapt these assorted proposals into a uniform framework
and provide a fair comparison of their results,
a large amount of underlying conceptual work was also required.
For instance, almost all of the proposed enhancements were
designed for systems that
perform the occurs-check and some of them were developed for
rather different abstract domains:
besides changing the representation of the domain elements,
such a situation usually requires a reconsideration of
the specification of the abstract operators.

All the experiments have been conducted using the
\china{} analyzer~\cite{Bagnara97th}
on a GNU/Linux PC system equipped with
an AMD Athlon clocked at 700 MHz and 256 MB of RAM.
\china{} is a data-flow analyzer for \CLPHofN{} languages
(i.e., ISO Prolog, \CLPR{}, \texttt{clp(FD)} and so forth),
$\HofN$ being an extended Herbrand system where the values of
a numeric domain $\cN$ can occur as leaves of the terms.
\china{}, which is written in \Cplusplus{},
performs bottom-up analysis deriving information on both
call-patterns and success-patterns by means of program transformations
and optimized fixpoint computation techniques.
An abstract description is computed for the call- and success-patterns
for each predicate defined in the program using a sophisticated chaotic
iteration strategy proposed in
\cite{Bourdoncle93,Bourdoncle93th}.\footnote{\china{} uses the recursive
fixpoint iteration strategy on the weak topological ordering defined
by partitioning of the call graph into strongly-connected subcomponents
\cite{Bourdoncle93th}.}

A major point of the experimental evaluation is given
by the test-suite, which is probably the largest one ever reported
in the literature on data-flow analysis of (constraint) logic programs.
The suite comprises all the programs we have access to
(i.e., everything we could find by systematically dredging the Internet):
more than 330 programs, 24 MB of code, 800 K lines.
Besides classical benchmarks,
several real programs of respectable size are included,
the largest one containing 10063 clauses in 45658 lines of code.
The suite also comprises a few synthetic benchmarks,
which are artificial programs explicitly constructed to stress
the capabilities of the analyzer and of its abstract domains
with respect to precision and/or efficiency.

Because of the exponential complexity of the base domain $\SFL$,
a data-flow analysis that includes this domain
will only be practical if it incorporates widening
operators~\cite{ZaffanellaBH99}.\footnote{Note that we use the term
`widening operator' in its broadest sense: any mechanism whereby,
in the course of the analysis, an abstract description is substituted
by one that is less precise.}
However, since almost none of the investigated combinations
come with specialized widening operators,
for a fair assessment of the precision improvements
we decided to disable all the widenings available
in our $\SFL$ implementation.
As a consequence,
there are a few benchmarks for which the analysis does not terminate
in reasonable time or absorbs memory beyond acceptable limits,
so that a precision comparison is not possible.
Note however that the motivations behind this choice go beyond
the simple observation that widening operators affect the precision
of the analysis: the problem is also that, if we use the widenings
defined and tuned for our implementation of the domain $\SFL$,
the results would be biased.
In fact, the definition of a good widening for an analysis
domain normally depends on both the representation and the implementation
of the domain.
In other words, different implementations even of the same domain
will require different tunings of the widening operators
(or even, possibly, brand new widenings).
This means that adopting the same widening operators
for all the domain combinations
would weaken, if not invalidate,
any conclusions regarding the relative benefits
of the investigated enhancements.
On the other hand, the definition of a new specialized
widening operator for each one of the considered domain combinations,
besides being a formidable task,
would also be wasted effort as the number of benchmark programs
for which termination cannot be obtained within reasonable time is really
small.

For space reasons, the experimental results
are only summarized here.
The interested reader can find more information
(including a description of the constantly growing benchmark suite
and detailed results for each benchmark)
at the URI \texttt{http://www.cs.unipr.it/China/}.
Indeed, given the high number of benchmark programs
and the many domain combinations considered,\footnote{We compute
the results of 40 different variations of the static analysis,
which are then used to perform 36 comparisons.
The results are computed over 332 programs for goal-independent analyses
and over 221 programs for goal-dependent analyses.
This difference in the number of benchmarks considered
comes from the fact that many programs
either are not provided with a set of entry goals
or use constructs such as \texttt{call(G)}
where \texttt{G} is a term whose principal functor is not known.
In these cases the analyzer recognizes that goal-dependent analysis
is pointless, since no call-patterns can be excluded.}
even finding a concise, meaningful and practical way to summarize
the results has been a non-trivial task.

For each benchmark,
precision is measured by counting the number of independent pairs
(the corresponding columns are labeled `I' in the tables)
as well as the numbers of definitely ground (labeled `G'), free (`F')
and linear (`L') variables detected by each abstract domain.
The results obtained for different analyses are compared
by computing the relative precision improvements or degradations
on each of these quantities and expressing them using percentages.
The ``overall'' (`O') precision improvement for the benchmark
is also computed as the maximum improvement
on all the measured quantities.\footnote{When computing
this ``overall'' result for a benchmark, the presence of even
a single precision loss for one of the measures overrides
any precision improvement
computed on the other components.}
The benchmark suite is then partitioned into several
precision equivalence classes: the cardinalities of these classes
are expressed again using percentages.
For example, when looking at the precision results
reported in Table~\ref{tab:sfl-vs-simplepos}
for goal-dependent analysis,
the value 2.3 that can be found at the intersection
of the row labeled `$0 < p \leq 2$' with the column labeled `G'
is to be read as follows:
``for 2.3 percent of the benchmarks the increase in
the number of ground variables is less than or equal to 2 percent.''
The precision class labeled `unknown' identifies
those benchmarks for which a precision comparison was not possible,
because one or both of the analyses was timed-out
(for all comparisons, the time-out threshold is 600 seconds).
In summary, a precision table gives an approximation of the distribution
of the programs in the benchmark suite with respect to the obtained
precision gains.

For a rough estimate of the efficiency
of the different analyses, for each comparison we provide
two tables that summarize the times taken by the fixpoint computations.
It should be stressed that these by no means provide a faithful account
of the intrinsic computational cost of the tested domain combinations.
Besides the lack of widenings, which have a big impact on performance
as can be observed by the results reported in \cite{ZaffanellaBH99},
the reader should not forget that, for ease of implementation,
having targeted at precision we traded efficiency whenever possible.
Therefore, these tables provide, so to speak, upper-bounds:
refined implementations can be expected to perform at least as well as
those reported in the tables.

As done for the precision results, the timings are summarized
by partitioning the suite into equivalence classes
and reporting the cardinality of each class using percentages.
In the first table we consider the distribution
of the \emph{absolute time differences},
that is we measure the slow-down and speed-up
due to the incorporation of the considered enhancement.
Note that the class called `same time' actually comprises the benchmarks
having a time difference below a given threshold,
which is fixed at 0.1 seconds.
In the second table we show the distribution of the
\emph{total fixpoint computation times},
both for the base analysis (in the columns labeled `\%1')
and for the enhanced one (in the columns labeled `\%2');
the columns labeled `$\Delta$' show how much each total time class
grows or shrinks due to the inclusion of the considered combination.

\section{A Simple Combination with Pos}
\label{sec:simplepos}

It is well-known that the domain $\Sharing$ (and thus also $\SFL$)
keeps track of ground dependencies.
More precisely, $\Sharing$ contains $\Def$, the domain of definite
Boolean functions \cite{ArmstrongMSS98},
as a proper subdomain \cite{CortesiFW92,ZaffanellaHB99}.
However, we consider here the combination of $\SFL$ with $\Pos$,
the domain of positive Boolean functions \cite{ArmstrongMSS98}.
There are several good reasons to couple $\SFL$ with $\Pos$:
\begin{enumerate}
\item
$\Pos$ is strictly more expressive than $\Def$ in that
it can represent (positive) disjunctive groundness dependencies
that arise in the analysis of Prolog
programs \cite{ArmstrongMSS98}.
The ability to deal with disjunctive dependencies is also needed
for the precise approximation of the constraints
of some CLP languages:
for example, when using the finite domain solver of SICStus Prolog,
the user can write disjunctive constraints such as
`\verb+X #= 4 #\/ Y #= 6+'.
\item
\label{benefit-precision}
The increased precision on groundness propagates to the $\SFL$ component.
It can be exploited to remove redundant sharing groups and
to identify more linear variables, therefore having a positive impact
on the computation of the $\amgu$ operator of the $\SFL$ domain.
Moreover, when dealing with sequences of bindings, the added
groundness information allows them to be usefully reordered.
In fact, while it has been proved that $\Sharing$ alone is commutative,
meaning that the result of the analysis does not depend on the ordering
in which the bindings are executed \cite{HillBZ98b},
the domain $\SFL$ does not enjoy this property.
In particular, even for the simpler combination of $\Sharing$
with linearity it is known since \cite[pp.~66-67]{Langen90th}
that better results are obtained if the \emph{grounding bindings}
are considered before the others.\footnote{A binding
$x=t$ is \emph{grounding} with respect to an abstract description if,
in all the concrete computation states approximated by the abstract
description, either the variable $x$ is ground or all the variables in $t$
are ground.
For example, when considering an abstract description $\sh \in \SH$,
the binding $x = t$ is grounding if $\rel(\{x\}, \sh) = \emptyset$ or
$\rel(\vars(t), \sh) = \emptyset$.}
As an example, consider the sequences of unifications
$\bigl(f(X, X, Y) = A,\, X = a\bigr)$ and
$\bigl(X = a,\, f(X, X, Y) = A\bigr)$ \cite[p.~66]{Langen90th}.
The combination with $\Pos$ is clearly advantageous in this respect.
\item
Besides being useful for improving precision on other properties,
disjunctive dependencies also have a few direct applications,
such as occurs-check reduction.
As observed in~\cite{CrnogoracKS96}, if the groundness formula
$x \lor y$ holds, the unification $x = y$ is occurs-check free,
even when neither $x$ nor $y$ are definitely linear.
\item
\label{benefit-efficiency}
Detecting the set of definitely ground variables through $\Pos$
and exploiting it to simplify the operations on $\SFL$
can improve the efficiency of the analysis.
In particular this is true if the set of ground variables
is readily available, as is the case, for instance,
with the GER implementation of $\Pos$ \cite{BagnaraS99}.
\item
The combination with $\Pos$ is essential for the application
of a powerful widening technique on $\SFL$ \cite{ZaffanellaBH99}.
This is very important, since analysis based on $\SFL$ is not practical
without widenings.
\item
In the context of the analysis of CLP programs, the notions
of ``ground variable'' and the notion of ``variable that cannot share
a common variable with other variables'' are distinct.
A numeric variable in, say, \CLPR{}, cannot share with other numerical
variables (not in the sense of interest in this paper)
but is not ground unless it has been constrained to a unique value.
Thus the analysis of CLP programs with $\SFL$ alone either will lose
precision on pair-sharing (if arithmetic constraints are abstracted
into ``sharings'' among numeric variables in order to approximate
the groundness of the latter)
or will be imprecise on the groundness of numeric variables
(because only Herbrand constraints take part in the construction of
sharing-sets).
In the first alternative, as we have already noted, the precision
with which groundness of numeric variables can be tracked will
also be limited.
Since groundness of numeric variables is important for a number
of applications (e.g., compiling equality constraints down to
assignments or tests in some circumstances), we advocate the use
of $\Pos$ and $\SFL$ at the same time.

\end{enumerate}

\begin{table*}
\centering
\begin{tabular}{||c||r||r|r|r|r||r||r|r|r|r||}
\hhline{|t:=:t:=====:t:=====:t|}
 Prec. class
  & \multicolumn{5}{c||}{Goal Independent}
  & \multicolumn{5}{c||}{Goal Dependent} \\
\hhline{|b:=::=:t:====::=:t:====:|}
 \multicolumn{1}{c||}{}
 & \multicolumn{1}{c||}{O}
 & \multicolumn{1}{c|}{I}
 & \multicolumn{1}{c|}{G}
 & \multicolumn{1}{c|}{F}
 & \multicolumn{1}{c||}{L}
 & \multicolumn{1}{c||}{O}
 & \multicolumn{1}{c|}{I}
 & \multicolumn{1}{c|}{G}
 & \multicolumn{1}{c|}{F}
 & \multicolumn{1}{c||}{L} \\
\hhline{|t:=::=::====::=::====:|}
$\phantom{-1}5 < p \leq 10\phantom{-}$ & --- & --- & --- & --- & --- & 0.5 & --- & 0.5 & --- & --- \\
\hhline{||-||-||-|-|-|-||-||-|-|-|-||}
$\phantom{-1}2 < p \leq 5\phantom{-0}$ & 0.3 & --- & 0.3 & --- & --- & --- & --- & --- & --- & --- \\
\hhline{||-||-||-|-|-|-||-||-|-|-|-||}
$\phantom{-1}0 < p \leq 2\phantom{-0}$ & 0.6 & 0.6 & 0.6 & --- & 0.6 & 3.2 & 3.6 & 2.3 & --- & 2.7 \\
\hhline{|:=::=::=:=:=:=::=::=:=:=:=:|}
same precision & 95.8 & 96.1 & 95.8 & 96.7 & 96.1 & 92.8 & 92.8 & 93.7 & 96.4 & 93.7 \\
\hhline{||-||-||-|-|-|-||-||-|-|-|-||}
unknown & 3.3 & 3.3 & 3.3 & 3.3 & 3.3 & 3.6 & 3.6 & 3.6 & 3.6 & 3.6 \\
\hhline{|b:=:b:=:b:====:b:=:b:====:b|}
\end{tabular}

\bigskip\bigskip

\begin{tabular}{||l||r|r||}
\hhline{|t:=:t:==:t|}
\multicolumn{1}{||c||}{Time difference class} &
\multicolumn{2}{c||}{\% benchmarks} \\
\hhline{|b:=::==:|}
\multicolumn{1}{c||}{} &
\multicolumn{1}{c|}{Goal Ind.} &
\multicolumn{1}{c||}{Goal Dep.} \\
\hhline{|t:=::==:|}
$\phantom{0.5 < \mathord{}} \text{degradation} > 1$ & 2.7 & 6.8 \\
 \hhline{||-||-|-||}
$0.5 < \text{degradation} \leq 1$ & 1.5 & 0.5 \\
 \hhline{||-||-|-||}
$0.2 < \text{degradation} \leq 0.5$ & 3.0 & 0.9 \\
 \hhline{||-||-|-||}
$0.1 < \text{degradation} \leq 0.2$ & 5.7 & 5.0 \\
 \hhline{|:=::=:=:|}
$\phantom{0.5 < \mathord{}} \text{both timed out}$ & 3.3 & 3.6 \\
 \hhline{||-||-|-||}
$\phantom{0.5 < \mathord{}} \text{same time}$ & 81.6 & 81.9 \\
 \hhline{|:=::=:=:|}
$0.1 < \text{improvement} \leq 0.2$ & --- & 0.5 \\
 \hhline{||-||-|-||}
$0.2 < \text{improvement} \leq 0.5$ & 0.9 & 0.5 \\
 \hhline{||-||-|-||}
$0.5 < \text{improvement} \leq 1$ & 0.3 & --- \\
 \hhline{||-||-|-||}
$\phantom{0.5 < \mathord{}} \text{improvement} > 1$ & 0.9 & 0.5 \\
\hhline{|b:=:b:==:b|}
\end{tabular}

\bigskip\bigskip

\begin{tabular}{||c||r|r|r||r|r|r||}
\hhline{|t:=:t:===:t:===:t|}
\multicolumn{1}{||c||}{Total time class}
 & \multicolumn{3}{c||}{Goal Ind.}
 & \multicolumn{3}{c||}{Goal Dep.} \\
\hhline{|b:=::===::===:|}
\multicolumn{1}{c||}{} & \multicolumn{1}{c|}{\%1}
 & \multicolumn{1}{c|}{\%2} & \multicolumn{1}{c||}{$\Delta$}
 & \multicolumn{1}{c|}{\%1} & \multicolumn{1}{c|}{\%2}
 & \multicolumn{1}{c||}{$\Delta$} \\
\hhline{|t:=::===::===:|}
$\text{timed out}$ & 3.3 & 3.3 & --- & 3.6 & 3.6 & --- \\
 \hhline{||-||-|-|-||-|-|-||}
$\phantom{0.5 < \mathord{}} t > 10\phantom{.}$ & 8.4 & 9.0 & 0.6 & 7.2 & 7.2 & --- \\
 \hhline{||-||-|-|-||-|-|-||}
$\phantom{0.}5 < t \leq 10\phantom{.}$ & 0.6 & 0.3 & -0.3 & 1.4 & 1.4 & --- \\
 \hhline{||-||-|-|-||-|-|-||}
$\phantom{0.}1 < t \leq 5\phantom{.5}$ & 6.6 & 7.5 & 0.9 & 3.2 & 3.6 & 0.5 \\
 \hhline{||-||-|-|-||-|-|-||}
$0.5 < t \leq 1\phantom{.5}$ & 3.3 & 2.7 & -0.6 & 5.4 & 5.4 & --- \\
 \hhline{||-||-|-|-||-|-|-||}
$0.2 < t \leq 0.5$ & 7.2 & 8.4 & 1.2 & 10.4 & 13.1 & 2.7 \\
 \hhline{||-||-|-|-||-|-|-||}
$\phantom{0.5 < \mathord{}} t \leq 0.2$ & 70.5 & 68.7 & -1.8 & 68.8 & 65.6 & -3.2 \\
\hhline{|b:=:b:===:b:===:b|}
\end{tabular}

\caption{$\PSDFL$ versus $\Pos \times \PSDFL$.}
\label{tab:sfl-vs-simplepos}
\end{table*}

Thus, as a first technique to enhance the precision of sharing analysis,
we consider the simple propagation
of the set of definitely ground variables from the $\Pos$ component
to the $\SFL$ component.\footnote{A more precise
combination will be considered in Section~\ref{sec:enhancedpos}.}
We denote this domain by $\Pos \times \SFL$.

As noted above, the GER implementation of \cite{BagnaraS99},
besides being the fastest implementation of $\Pos$ known to date,
is the natural candidate for this combination, since it provides
constant-time access to the set $G$ of the definitely ground variables.
Note that the widenings on the $\Pos$ component have been retained.
The reason for this choice is that they fire for only a few benchmarks
and, when coming into play, they rarely affect the precision
of the groundness analysis:
by switching them off we would only obtain a few more time-outs.

In the $\SFL$ component, the set $G$ of definitely ground variables
is used
\begin{itemize}
\item
to reorder the sequence of bindings in the abstract unification
so as to handle the grounding ones first;
\item
to eliminate the sharing groups containing
at least one ground variable; and
\item
to recover from previous linearity losses.
\end{itemize}

The experimental results for $\Pos \times \SFL$ are compared
with those obtained for the domain $\SFL$ considered in isolation
and reported in Table~\ref{tab:sfl-vs-simplepos}.
It can be observed that a precision improvement is observed
in all of the measured quantities but freeness,
affecting up to 3.6\% of the programs.

Note that there is a small discrepancy between these results
and those of~\cite{BagnaraZH00}
where more improvements were reported.
The reason is that the current $\SFL$ implementation uses
an enhanced abstract unification operator,
fully exploiting the anticipation of the grounding bindings
even on the base domain $\SFL$ itself.
In contrast, in the earlier $\SFL$ implementation used for the
results in~\cite{BagnaraZH00},
only the \emph{syntactically} grounding bindings
were anticipated.\footnote{A binding $x = t$ is syntactically grounding
if $\vars(t) = \emptyset$. This ``syntactic'' definition differs from
the ``semantic'' one provided before in that it does not depend on
the information provided by an abstract description.}

As for the timings, even if the figures in the tables
seem to contradict what we claimed in point~\ref{benefit-efficiency} above,
a closer inspection of the detailed results reveals that
this is only due to a very unfortunate interaction between
the increased precision given by $\Pos$
and the absence of widening operators on $\SFL$.
This state of affairs forces the analyzer to compute a few,
but very expensive, further iterations in the fixpoint computation.

Because of the reasons detailed above,
we believe $\Pos$ should be part of the global domain
employed by any ``production analyzer'' for CLP languages.
That is why, for the remaining comparisons, unless otherwise stated,
this simple combination with the $\Pos$ domain
is always included.

\section{Tracking Explicit Structural Information}
\label{sec:struct}

A way of increasing the precision of almost any analysis domain
is by enhancing it with structural information.
For mode analysis, this idea dates back to \cite{JanssensB92}.
A more general technique was proposed in~\cite{CortesiLCVH94},
where the generic structural domain $\texttt{Pat}(\Re)$ was introduced.
A similar proposal, tailored to sharing analysis, is due to
\cite{BruynoogheCM94}, where \emph{abstract equation systems} are considered.
In the experimental evaluation the  $\Pattern(\cdot)$ construction
\cite{Bagnara97th,Bagnara97,BagnaraHZ00} is used.
This is similar to $\texttt{Pat}(\Re)$ and correctly supports the
analysis of languages omitting the occurs-check in the unification procedure
as well as those that do not.

The construction $\Pattern(\cdot)$ upgrades a domain $\cD$
(which must support a certain set of basic operations)
with structural information. The resulting domain,
where structural information is retained to some extent,
is usually much more precise than $\cD$ alone.
There are many occasions where these precision gains
give rise to consistent speed-ups.
The reason for this is twofold.
First, structural information has the potential of pruning
some computation paths on the grounds that they cannot be followed
by the program being analyzed.
Second, maintaining a tuple of terms with many variables,
each with its own description, can be cheaper than
computing a description for the whole tuple \cite{BagnaraHZ00}.
Of course, there is also a price to be paid:
in the analysis based on $\Pattern(\cD)$,
the elements of $\cD$ that are to be manipulated are often bigger
(i.e., there are more variables of interest)
than those that arise in analyses that are simply based on $\cD$.

When comparing the precision results, the difference in the number
of variables tracked by the two analyses poses a non-trivial problem.
How can we provide a \emph{fair} measure of the precision gain?
There is no easy answer to such a question.
The approach chosen is simple though unsatisfactory:
at the end of the analysis, first throw away all the structural
information in the results and then calculate the cardinality of
the usual sets.
In other words, we only measure how the explicit structural information
in $\Pattern(\cD)$ improves the precision on $\cD$ itself,
which is only a tiny part of the real gain in accuracy.
As shown by the following example, this solution greatly
underestimates the precision improvement coming from
the integration of structural information.

Consider a simple but not trivial Prolog program:
\texttt{mastermind}.\footnote{This program which implements
the game ``Mastermind'' was rewritten by H.~Koenig and T.~Hoppe
after code by M.~H.~van Emden
and available at
\url{http://www.cs.unipr.it/China/Benchmarks/Prolog/mastermind.pl}.}
Consider also the only direct query for which it has been written,
`\texttt{?- play.}', and focus the attention
on the procedure \texttt{extend\_code/1}.
A standard goal-dependent analysis of the program
with the $\Pos \times \SFL$ domain cannot say anything
on the successes of \texttt{extend\_code/1}.
If the analysis is performed with $\Pattern(\Pos \times \SFL)$
the situation changes radically.
Here is what such a domain allows \china{} to derive:\footnote{Some
extra groundness information obtained by the analysis
has been omitted for simplicity:
this says that, if \texttt{A} and \texttt{B}
turn out to be ground, then \texttt{E} will also be ground.}
\begin{verbatim}
extend_code([([A|B],C,D)|E]) :-
  list(B), list(E),
  (functor(C,_,1);integer(C)),
  (functor(D,_,1);integer(D)),
  ground([C,D]), may_share([[A,B,E]]).
\end{verbatim}
This means: ``during any execution of the program,
whenever \texttt{extend\_code/1} succeeds it will have
its argument bound
to a term of the form \texttt{[([A|B],C,D)|E]},
where
\texttt{B} and \texttt{E} are bound to list cells
(i.e., to terms whose principal
functor is either \texttt{'.'/2} or \texttt{[]/0});
\texttt{C} and \texttt{D} are ground
and bound to a functor of arity 1 or to an integer;
and pair-sharing may only occur among
\texttt{A},  \texttt{B}, and \texttt{E}''.
Once structural information has been discarded,
the analysis with $\Pattern(\Pos \times \SFL)$ only specifies that
\texttt{extend\_code/1} may succeed.
Thus, according to our approach to the precision comparison,
explicit structural information gives no improvements in the analysis
of \texttt{extend\_code/1}
(which is far from being a fair conclusion).

Of course, structural information is very valuable in itself.
For example, when exploited for optimized compilation it allows
for enhanced clause indexing and simplified unification.
Several other semantics-based program manipulation techniques
(such as debugging, program specialization, and verification)
benefit from this kind of information.
However, the value of this extra precision could only be measured
from the point of view of the target application of the analysis.

\begin{table*}
\centering
\begin{tabular}{||c||r||r|r|r|r||r||r|r|r|r||}
\hhline{|t:=:t:=====:t:=====:t|}
 Prec. class
  & \multicolumn{5}{c||}{Goal Independent}
  & \multicolumn{5}{c||}{Goal Dependent} \\
\hhline{|b:=::=:t:====::=:t:====:|}
 \multicolumn{1}{c||}{}
 & \multicolumn{1}{c||}{O}
 & \multicolumn{1}{c|}{I}
 & \multicolumn{1}{c|}{G}
 & \multicolumn{1}{c|}{F}
 & \multicolumn{1}{c||}{L}
 & \multicolumn{1}{c||}{O}
 & \multicolumn{1}{c|}{I}
 & \multicolumn{1}{c|}{G}
 & \multicolumn{1}{c|}{F}
 & \multicolumn{1}{c||}{L} \\
\hhline{|t:=::=::====::=::====:|}
$\phantom{-20 < \mathord{}} p > 20\phantom{-}$ & 7.5 & 2.7 & 3.9 & 2.1 & 3.3 & 6.3 & 1.4 & 3.6 & 1.8 & 3.6 \\
\hhline{||-||-||-|-|-|-||-||-|-|-|-||}
$\phantom{-}10 < p \leq 20\phantom{-}$ & 3.9 & 2.1 & 2.7 & --- & 2.4 & 2.7 & 2.3 & 1.4 & --- & 2.7 \\
\hhline{||-||-||-|-|-|-||-||-|-|-|-||}
$\phantom{-1}5 < p \leq 10\phantom{-}$ & 4.5 & 1.8 & 2.7 & 2.4 & 2.4 & 1.8 & 0.9 & 2.3 & 0.9 & 1.4 \\
\hhline{||-||-||-|-|-|-||-||-|-|-|-||}
$\phantom{-1}2 < p \leq 5\phantom{-0}$ & 7.5 & 6.0 & 3.9 & 2.7 & 5.1 & 2.7 & 3.2 & 1.4 & 1.8 & 2.3 \\
\hhline{||-||-||-|-|-|-||-||-|-|-|-||}
$\phantom{-1}0 < p \leq 2\phantom{-0}$ & 7.8 & 9.0 & 6.6 & 6.9 & 12.0 & 2.3 & 4.5 & 1.8 & 1.8 & 5.0 \\
\hhline{|:=::=::=:=:=:=::=::=:=:=:=:|}
same precision & 61.7 & 71.7 & 73.5 & 79.2 & 67.8 & 74.2 & 78.3 & 80.1 & 84.2 & 75.1 \\
\hhline{||-||-||-|-|-|-||-||-|-|-|-||}
unknown & 6.6 & 6.6 & 6.6 & 6.6 & 6.6 & 9.5 & 9.5 & 9.5 & 9.5 & 9.5 \\
\hhline{|:=::=::=:=:=:=::=::=:=:=:=:|}
$\phantom{-10 < \mathord{}} p < 0\phantom{-0}$ & 0.3 & --- & --- & --- & 0.3 & 0.5 & --- & --- & --- & 0.5 \\
\hhline{|b:=:b:=:b:====:b:=:b:====:b|}
\end{tabular}

\bigskip\bigskip

\begin{tabular}{||l||r|r||}
\hhline{|t:=:t:==:t|}
\multicolumn{1}{||c||}{Time diff. class} &
\multicolumn{2}{c||}{\% benchmarks} \\
\hhline{|b:=::==:|}
\multicolumn{1}{c||}{} &
\multicolumn{1}{c|}{Goal Ind.} &
\multicolumn{1}{c||}{Goal Dep.} \\
\hhline{|t:=::==:|}
$\phantom{0.5 < \mathord{}} \text{degradation} > 1$ & 11.7 & 17.6 \\
 \hhline{||-||-|-||}
$0.5 < \text{degradation} \leq 1$ & 1.2 & 0.9 \\
 \hhline{||-||-|-||}
$0.2 < \text{degradation} \leq 0.5$ & 3.6 & 4.1 \\
 \hhline{||-||-|-||}
$0.1 < \text{degradation} \leq 0.2$ & 1.5 & 4.1 \\
 \hhline{|:=::=:=:|}
$\phantom{0.5 < \mathord{}} \text{both timed out}$ & 3.3 & 3.6 \\
 \hhline{||-||-|-||}
$\phantom{0.5 < \mathord{}} \text{same time}$ & 70.8 & 66.5 \\
 \hhline{|:=::=:=:|}
$0.1 < \text{improvement} \leq 0.2$ & 0.9 & 0.5 \\
 \hhline{||-||-|-||}
$0.2 < \text{improvement} \leq 0.5$ & 1.5 & --- \\
 \hhline{||-||-|-||}
$0.5 < \text{improvement} \leq 1$ & 0.6 & 0.5 \\
 \hhline{||-||-|-||}
$\phantom{0.5 < \mathord{}} \text{improvement} > 1$ & 4.8 & 2.3 \\
\hhline{|b:=:b:==:b|}
\end{tabular}

\bigskip\bigskip

\begin{tabular}{||c||r|r|r||r|r|r||}
\hhline{|t:=:t:===:t:===:t|}
\multicolumn{1}{||c||}{Total time class}
 & \multicolumn{3}{c||}{Goal Ind.}
 & \multicolumn{3}{c||}{Goal Dep.} \\
\hhline{|b:=::===::===:|}
\multicolumn{1}{c||}{} & \multicolumn{1}{c|}{\%1}
 & \multicolumn{1}{c|}{\%2} & \multicolumn{1}{c||}{$\Delta$}
 & \multicolumn{1}{c|}{\%1} & \multicolumn{1}{c|}{\%2}
 & \multicolumn{1}{c||}{$\Delta$} \\
\hhline{|t:=::===::===:|}
$\text{timed out}$ & 3.3 & 6.6 & 3.3 & 3.6 & 9.5 & 5.9 \\
 \hhline{||-||-|-|-||-|-|-||}
$\phantom{0.5 < \mathord{}} t > 10\phantom{.}$ & 9.0 & 8.4 & -0.6 & 7.2 & 8.6 & 1.4 \\
 \hhline{||-||-|-|-||-|-|-||}
$\phantom{0.}5 < t \leq 10\phantom{.}$ & 0.3 & 1.5 & 1.2 & 1.4 & 1.8 & 0.5 \\
 \hhline{||-||-|-|-||-|-|-||}
$\phantom{0.}1 < t \leq 5\phantom{.5}$ & 7.5 & 6.6 & -0.9 & 3.6 & 5.0 & 1.4 \\
 \hhline{||-||-|-|-||-|-|-||}
$0.5 < t \leq 1\phantom{.5}$ & 2.7 & 3.3 & 0.6 & 5.4 & 3.2 & -2.3 \\
 \hhline{||-||-|-|-||-|-|-||}
$0.2 < t \leq 0.5$ & 8.4 & 10.2 & 1.8 & 13.1 & 13.6 & 0.5 \\
 \hhline{||-||-|-|-||-|-|-||}
$\phantom{0.5 < \mathord{}} t \leq 0.2$ & 68.7 & 63.3 & -5.4 & 65.6 & 58.4 & -7.2 \\
\hhline{|b:=:b:===:b:===:b|}
\end{tabular}
\caption{$\Pos \times \PSDFL$ versus $\Pattern(\Pos \times \PSDFL)$.}
\label{tab:nostruct-vs-struct}
\end{table*}

Thus the precision of the domain $\Pos \times \SFL$ has been compared
with that obtained using the domain $\Pattern(\Pos \times \SFL)$ and
the results reported in Table~\ref{tab:nostruct-vs-struct}.
It can be seen that, for goal-independent analysis, on one third
of the benchmarks compared there is a precision improvement
in at least one of the measured quantities;
the same happens for one sixth of the benchmarks in the case of
goal-dependent analysis.
Moreover, the increase in precision can be considerable, as testified by
the percentages of benchmarks falling in the higher precision classes.

The reader may be surprised, as the authors were, to see that in
some cases the precision actually decreased.\footnote{This happens
for the program \texttt{attractions2}
in the case of goal-independent analysis and
for the program \texttt{semi} in the case of goal-dependent analysis.}
Indeed, to the best of our knowledge, this possibility has escaped
all previous research work investigating this kind of abstract domain
enhancement, including~\cite{CortesiLCVH94,BruynoogheCM94,Bagnara97th}.
The reason for these precision losses lies in a subtle interaction
between the explicit structural information and
the underlying abstract unification operator.

When using the base domain $\Pos \times \SFL$,
the abstract evaluation of a single syntactic binding,
such as $x = f(y, z)$,
directly corresponds to a single application of the $\amgu$ operator.
In contrast, when computing on $\Pattern(\Pos \times \SFL)$,
it may well happen that the computed abstract description already
contains the information that variable $x$ is bound to a term,
such as $f\bigl(g(w), w\bigr)$.
As a consequence, after peeling the principal functor $f/2$,
the abstract computation should proceed by evaluating,
on the base domain $\Pos \times \SFL$,
the set of bindings $\bigl\{ y = g(w), z = w \bigr\}$.
Here the problem is that, as already noted, the $\amgu$ operator
on the base domain $\Pos \times \SFL$ is not commutative.
While this improvement in the data
 used by the abstract computation
very often allows for a corresponding increase in the precision
of the result,
in rare situations it may happen that a sub-optimal ordering
of the bindings is chosen, incurring a precision loss.

It should be noted that such a negative interaction with
the explicit structural information is only possible when
the underlying domain implements non-commutative abstract operators.
In particular, this phenomenon could not be observed when
computing on $\Pattern(\SH)$ or $\Pattern(\Pos)$.

One issue that should be resolved is
 whether the improvements provided
by explicit structural information subsume those previously obtained
for the simple combination with $\Pos$.
Intuitively, it would seem that this cannot happen,
since these two enhancements
are based on different kinds of information:
while the $\Pattern(\cdot)$ construction encodes some
\emph{definite} structural information, the precision gain
due to using $\Pos$ rather than just $\Def$ only stems from
\emph{disjunctive} groundness dependencies.
However, the impact of these techniques on the overall analysis
is really intricate and some overlapping cannot be excluded
\emph{a priori}: for instance, both techniques affect the ordering of bindings
in the computation of abstract unification on $\SFL$.
In order to provide some experimental evidence for this qualitative
reasoning, the precision results are computed for the simpler domain
$\Pattern(\SFL)$ and then compared with those obtained for the
domain $\Pattern(\Pos \times \SFL)$.
Since the main differences between
Tables~\ref{tab:sfl-vs-simplepos}
and~\ref{tab:sfl-struct-vs-simplepos-struct}
can be explained by discrepancies in the numbers of programs that timed-out,
these results confirm our expectations
that these two enhancements are effectively orthogonal.

\begin{table*}
\centering
\begin{tabular}{||c||r||r|r|r|r||r||r|r|r|r||}
\hhline{|t:=:t:=====:t:=====:t|}
 Prec. class
  & \multicolumn{5}{c||}{Goal Independent}
  & \multicolumn{5}{c||}{Goal Dependent} \\
\hhline{|b:=::=:t:====::=:t:====:|}
 \multicolumn{1}{c||}{}
 & \multicolumn{1}{c||}{O}
 & \multicolumn{1}{c|}{I}
 & \multicolumn{1}{c|}{G}
 & \multicolumn{1}{c|}{F}
 & \multicolumn{1}{c||}{L}
 & \multicolumn{1}{c||}{O}
 & \multicolumn{1}{c|}{I}
 & \multicolumn{1}{c|}{G}
 & \multicolumn{1}{c|}{F}
 & \multicolumn{1}{c||}{L} \\
\hhline{|t:=::=::====::=::====:|}
$\phantom{-1}5 < p \leq 10\phantom{-}$ & --- & --- & --- & --- & --- & 0.5 & --- & 0.5 & --- & --- \\
\hhline{||-||-||-|-|-|-||-||-|-|-|-||}
$\phantom{-1}2 < p \leq 5\phantom{-0}$ & 0.3 & --- & 0.3 & --- & --- & --- & 0.5 & --- & --- & --- \\
\hhline{||-||-||-|-|-|-||-||-|-|-|-||}
$\phantom{-1}0 < p \leq 2\phantom{-0}$ & --- & --- & --- & --- & --- & 3.2 & 3.2 & 2.7 & --- & 2.7 \\
\hhline{|:=::=::=:=:=:=::=::=:=:=:=:|}
same precision & 93.1 & 93.4 & 93.1 & 93.4 & 93.4 & 86.4 & 86.4 & 86.9 & 90.0 & 87.3 \\
\hhline{||-||-||-|-|-|-||-||-|-|-|-||}
unknown & 6.6 & 6.6 & 6.6 & 6.6 & 6.6 & 10.0 & 10.0 & 10.0 & 10.0 & 10.0 \\
\hhline{|b:=:b:=:b:====:b:=:b:====:b|}
\end{tabular}

\bigskip\bigskip

\begin{tabular}{||l||r|r||}
\hhline{|t:=:t:==:t|}
\multicolumn{1}{||c||}{Time diff. class} &
\multicolumn{2}{c||}{\% benchmarks} \\
\hhline{|b:=::==:|}
\multicolumn{1}{c||}{} &
\multicolumn{1}{c|}{Goal Ind.} &
\multicolumn{1}{c||}{Goal Dep.} \\
\hhline{|t:=::==:|}
$\phantom{0.5 < \mathord{}} \text{degradation} > 1$ & 5.7 & 7.7 \\
 \hhline{||-||-|-||}
$0.5 < \text{degradation} \leq 1$ & 2.4 & 0.5 \\
 \hhline{||-||-|-||}
$0.2 < \text{degradation} \leq 0.5$ & 3.6 & 5.4 \\
 \hhline{||-||-|-||}
$0.1 < \text{degradation} \leq 0.2$ & 5.4 & 2.7 \\
 \hhline{|:=::=:=:|}
$\phantom{0.5 < \mathord{}} \text{both timed out}$ & 6.6 & 9.5 \\
 \hhline{||-||-|-||}
$\phantom{0.5 < \mathord{}} \text{same time}$ & 75.6 & 73.8 \\
 \hhline{|:=::=:=:|}
$0.1 < \text{improvement} \leq 0.2$ & --- & --- \\
 \hhline{||-||-|-||}
$0.2 < \text{improvement} \leq 0.5$ & 0.6 & --- \\
 \hhline{||-||-|-||}
$0.5 < \text{improvement} \leq 1$ & --- & --- \\
 \hhline{||-||-|-||}
$\phantom{0.5 < \mathord{}} \text{improvement} > 1$ & --- & 0.5 \\
\hhline{|b:=:b:==:b|}
\end{tabular}

\bigskip\bigskip

\begin{tabular}{||c||r|r|r||r|r|r||}
\hhline{|t:=:t:===:t:===:t|}
\multicolumn{1}{||c||}{Total time class}
 & \multicolumn{3}{c||}{Goal Ind.}
 & \multicolumn{3}{c||}{Goal Dep.} \\
\hhline{|b:=::===::===:|}
\multicolumn{1}{c||}{} & \multicolumn{1}{c|}{\%1}
 & \multicolumn{1}{c|}{\%2} & \multicolumn{1}{c||}{$\Delta$}
 & \multicolumn{1}{c|}{\%1} & \multicolumn{1}{c|}{\%2}
 & \multicolumn{1}{c||}{$\Delta$} \\
\hhline{|t:=::===::===:|}
$\text{timed out}$ & 6.6 & 6.6 & --- & 10.0 & 9.5 & -0.5 \\
 \hhline{||-||-|-|-||-|-|-||}
$\phantom{0.5 < \mathord{}} t > 10\phantom{.}$ & 8.1 & 8.4 & 0.3 & 7.7 & 8.6 & 0.9 \\
 \hhline{||-||-|-|-||-|-|-||}
$\phantom{0.}5 < t \leq 10\phantom{.}$ & 1.5 & 1.5 & --- & 2.3 & 1.8 & -0.5 \\
 \hhline{||-||-|-|-||-|-|-||}
$\phantom{0.}1 < t \leq 5\phantom{.5}$ & 5.1 & 6.6 & 1.5 & 4.5 & 5.0 & 0.5 \\
 \hhline{||-||-|-|-||-|-|-||}
$0.5 < t \leq 1\phantom{.5}$ & 3.9 & 3.3 & -0.6 & 3.2 & 3.2 & --- \\
 \hhline{||-||-|-|-||-|-|-||}
$0.2 < t \leq 0.5$ & 7.2 & 10.2 & 3.0 & 10.9 & 13.6 & 2.7 \\
 \hhline{||-||-|-|-||-|-|-||}
$\phantom{0.5 < \mathord{}} t \leq 0.2$ & 67.5 & 63.3 & -4.2 & 61.5 & 58.4 & -3.2 \\
\hhline{|b:=:b:===:b:===:b|}
\end{tabular}
\caption{$\Pattern(\PSDFL)$ versus $\Pattern(\Pos \times \PSDFL)$.}
\label{tab:sfl-struct-vs-simplepos-struct}
\end{table*}

Similar experimental evaluations, but based on the abstract
equation systems of \cite{BruynoogheCM94},
were reported by A.~Mulkers et al.~in~\cite{MulkersSJB94TR,MulkersSJB95}.
Here a depth-$k$ abstraction (replacing all
subterms occurring at a depth greater than or equal to $k$ with
fresh abstract variables) is conducted on a small benchmark suite
(19 programs) for values of $k$ between $0$ and $3$.
The domain they employed was not suitable for the analysis of real programs
and, in fact, even the analysis of a modest-sized program
like \texttt{ann}
could only be carried out with depth-$0$ abstraction
(i.e., without any structural information).
Such a problem in finding practical analyzers that incorporated
structural information with sharing analysis was not unique to this work:
there was at least one other previous attempt to evaluate the impact
of structural information on sharing analysis that failed because of
combinatorial explosion
[A.~Cortesi, personal communication, 1996].

What makes the more realistic experimentation now possible is the adoption
of the non-redundant domain $\PSD$, where the exponential star-union
operation is replaced by the quadratic self-bin-union.
Note that, even if biased by the absence of widenings,
the timings reported in Table~\ref{tab:nostruct-vs-struct}
show that the $\Pattern(\cdot)$ construction is computationally feasible.
Indeed, as demonstrated by the results reported in \cite{BagnaraHZ00},
an analyzer that incorporates a carefully designed structural information
component, besides being more precise, can also be very efficient.

The results obtained in this section demonstrate that
there is a relevant amount of sharing information
that is not detected when using the classical set-sharing domains.
Therefore, in order to provide an experimental evaluation
that is as systematic as possible,
in all of the remaining experiments the comparison is performed
both with and without explicit structural information.

\section{Reordering the Non-Grounding Bindings}
\label{sec:binding-ordering}

As already explained in Section~\ref{sec:simplepos},
 the results of abstract unification on $\SFL$
may depend on the order in which the bindings are considered
and will be improved
if the grounding bindings are considered first.
This heuristic, which has been used for all the experiments in this paper,
is well-known: in the literature all the examples that illustrate
the non-commutativity of the abstract $\mgu$ on $\SFL$ use
a grounding binding. However, as observed in Section~\ref{sec:struct},
the problem is more general than that.

To illustrate this, suppose that
$\VI = \{u,v,w,x,y,z\}$ is the set of relevant variables, and
consider the $\SFL$ element\footnote{Elements of $\SH$ are written
in a simplified notation, omitting the inner braces.
For instance, the set $\bigl\{ \{x\}, \{x,y\}, \{x,z\}, \{x,y,z\}\bigr\}$
is written as $\{ x, xy, xz, xyz \}$.}
\begin{align*}
  \sfl
    &\defeq \bigl\langle
             \{ vy, wy, xy, yz \},
             \emptyset,
             \{ u, x, z \}
           \bigr\rangle, \\
\intertext{%
where no variable is free and $u$, $x$, and $z$ are linear
with the bindings $v=w$ and $x=y$.
Then, applying $\amgu$ to these bindings in the given ordering, we have:
}
  \sfl_1
    &=
      \amgu( \sfl, v=w ) \\
        &=
          \bigl\langle
            \{ vwy, xy, yz \} ,
            \emptyset ,
            \{ u, x, z \}
          \bigr\rangle, \\
  \sfl_{1,2}
    &=
      \amgu( \sfl_1, x=y ) \\
        &=
          \bigl\langle
            \{ vwxy, vwxyz, xy, xyz \} ,
            \emptyset ,
            \{ u, z \}
          \bigr\rangle. \\
\intertext{%
Using the reverse ordering, we have:
}
  \sfl_2
    &=
      \amgu( \sfl, x=y ) \\
        &=
          \bigl\langle
            \{ vwxy, vwxyz, vxy, vxyz, wxy, wxyz, xy, xyz \},
              \emptyset,
              \{ u, z \}
          \bigr\rangle, \\
  \sfl_{2,1}
    &=
      \amgu( \sfl_2, v=w ) \\
        &=
          \bigl\langle
            \{ vwxy, vwxyz, xy, xyz \} ,
            \emptyset ,
            \{ u \}
          \bigr\rangle.
\end{align*}
Thus $\sfl_{2,1}$ loses the linearity of $z$
(which, in turn, could cause bigger precision losses
later in the analysis).

In principle, optimality can be obtained by adopting
the \emph{brute-force} approach:
trying all the possible orderings of the non-grounding bindings.
However, this is clearly not feasible.
While lacking a better alternative, it is reasonable to look for heuristics
that can be applied in the context of a \emph{local search} paradigm:
at each step, the next binding for the $\amgu$ procedure
is chosen by evaluating the effect of its abstract execution,
considered in isolation, on the precision of the analysis.

Suppose the number of independent pairs is taken as a measure of precision.
Then, at each step, for each of the bindings under consideration,
the new component $\sh'$,
as given by Definition~\ref{def:abs-funcs-SFL},
must be computed.
However, because the computation of $\sh'$ is
the most costly operation to be performed in the computation
of the $\amgu$ operator,
a direct application of this heuristic does not appear
to be feasible.
As an alternative, consider a heuristic based on
the number of star-unions that have to be computed.
Star-unions are likely to cause large losses in the number
of independent pairs that are found.
As only non-grounding bindings are considered,
any binding requiring the computation of a star-union will need
the star-union even if it is delayed,
although a binding that does not require the star-union
may require it
if its computation is postponed: its variables
may lose their freeness, linearity or independence as a result
of evaluating the other bindings.
It follows that one potential heuristic
is: ``delay the bindings requiring star-unions as much as possible''.
In the next example, by adopting this heuristic,
the linearity of variable $y$ is preserved.

Consider the application of the bindings $x = z$ and $v = w$
to the following abstract description:
\begin{align*}
  \sfl
    &\defeq
      \bigl\langle
        \{ vw, wx, wy, z \} ,
        \emptyset ,
        \{ u, v, x, y \}
      \bigr\rangle.
\end{align*}
Since $x$ is linear and independent from $z$,
computing $\amgu(\sfl, x=z)$ requires one star-union,
while two star-unions are needed when computing $\amgu(\sfl, v=w)$
because $v$ and $w$ may share.
Thus, with the proposed heuristic, $x=z$ is applied before $v=w$,
giving:
\begin{align*}
  \sfl_1
    &=
      \amgu( \sfl, x = z ) \\
        &=
          \bigl\langle
            \{ vw, wxz, wy \} ,
            \emptyset ,
            \{ u, v, y \}
          \bigr\rangle, \\
  \sfl_{1,2}
    &=
      \amgu( \sfl_1, v = w ) \\
        &=
          \bigl\langle
            \{ vw, vwxyz, vwxz, vwy \} ,
            \emptyset ,
            \{ u, y \}
          \bigr\rangle. \\
\intertext{%
In contrast, if $v=w$ is applied first, we have:
}
  \sfl_2
    &=
      \amgu( \sfl, v = w ) \\
        &=
          \bigl\langle
            \{ vw, vwx, vwxy, vwy, z \} ,
            \emptyset ,
            \{ u, x, y \}
          \bigr\rangle, \\
  \sfl_{2,1}
    &=
      \amgu( \sfl_2, x = z ) \\
        &=
          \bigl\langle
            \{ vw, vwxyz, vwxz, vwy \} ,
            \emptyset ,
            \{ u \}
          \bigr\rangle.
\end{align*}
Note that the same number of independent pairs
is computed in both cases.

It should be noted that this heuristic, considered in isolation,
is not a general solution and can actually lead to precision losses.
The problem is that, if a binding that needs a star-union is delayed,
then, when the star-union is computed, it may be done on a larger
sharing-set, forcing more (independent) pairs of variables
into the same sharing group.

Consider the application of the bindings $u = x$ and $v = w$
to the abstract description
\begin{align*}
  \sfl
    &\defeq
      \bigl\langle
        \{ u, uw, v, w, xy, xz \} ,
        \{ u, x \} ,
        \{ u, x \}
      \bigr\rangle. \\
\intertext{%
Since $x$ and $u$ are both free variables, no star-union
is needed in the computation of $\amgu(\sfl, u=x)$,
while two star-unions are needed when computing $\amgu(\sfl, v=w)$.
}
  \sfl_1
    &=
      \amgu( \sfl, u = x ) \\
        &=
          \bigl\langle
            \{ uwxy, uwxz, uxy, uxz, v, w \} ,
            \{ u, x \} ,
            \{ u, x \}
          \bigr\rangle, \\
  \sfl_{1,2}
    &=
      \amgu( \sfl_1, v = w ) \\
        &=
          \bigl\langle
            \{ uvwxy, uvwxyz, uvwxz, uxy, uxz, vw \} ,
            \emptyset,
            \emptyset
          \bigr\rangle. \\
\intertext{%
Using the other ordering, we have:
}
  \sfl_2
    &=
      \amgu( \sfl, v = w ) \\
        &=
          \bigl\langle
            \{ u, uvw, vw, xy, xz \} ,
            \{ x \} ,
            \{ x \}
          \bigr\rangle, \\
  \sfl_{2,1}
    &=
      \amgu( \sfl_2, u = x ) \\
        &=
          \bigl\langle
            \{
              uvwxy, uvwxz, uxy, uxz, vw \} ,
            \emptyset ,
            \emptyset
          \bigr\rangle.
\end{align*}
Note that in $\sfl_{2,1}$ variables $y$ and $z$ are independent,
whereas they may share in $\sfl_{1,2}$.
Thus, in this example,
by delaying the only binding that requires the star-unions, $v = w$,
the number of known independent pairs is decreased.

Another possibility is to consider a heuristic
that uses the numbers of free and linear variables
as a measure of precision for local optimization.
That is, it chooses first those bindings
for which these numbers are maximal.
However, the last example shown above is evidence that even such a proposal
may also cause precision losses
(the binding $u=x$ would be chosen first
as it preserves the freeness of variable $u$).

\begin{table*}
\centering
\begin{tabular}{||c||r||r|r|r|r||r||r|r|r|r||}
\hhline{~|t:=====:t:=====:t|}
    \multicolumn{1}{l||}{Goal Independent}
  & \multicolumn{5}{c||}{without Struct Info}
  & \multicolumn{5}{c||}{with Struct Info} \\
\hhline{|t:=::=:t:====::=:t:====:|}
 Prec. class
 & \multicolumn{1}{c||}{O}
 & \multicolumn{1}{c|}{I}
 & \multicolumn{1}{c|}{G}
 & \multicolumn{1}{c|}{F}
 & \multicolumn{1}{c||}{L}
 & \multicolumn{1}{c||}{O}
 & \multicolumn{1}{c|}{I}
 & \multicolumn{1}{c|}{G}
 & \multicolumn{1}{c|}{F}
 & \multicolumn{1}{c||}{L} \\
\hhline{|:=::=::====::=::====:|}
$\phantom{-1}0 < p \leq 2\phantom{-0}$ & 0.9 & --- & --- & --- & 0.9 & --- & --- & --- & --- & --- \\
\hhline{|:=::=::=:=:=:=::=::=:=:=:=:|}
same precision & 94.6 & 95.5 & 96.4 & 96.4 & 95.5 & 91.3 & 91.3 & 93.1 & 93.1 & 93.1 \\
\hhline{||-||-||-|-|-|-||-||-|-|-|-||}
unknown & 3.6 & 3.6 & 3.6 & 3.6 & 3.6 & 6.9 & 6.9 & 6.9 & 6.9 & 6.9 \\
\hhline{|:=::=::=:=:=:=::=::=:=:=:=:|}
$\phantom{1}-2 \leq p < 0\phantom{-2}$ & 0.9 & 0.9 & --- & --- & --- & 1.8 & 1.8 & --- & --- & --- \\
\hhline{|b:=:b:=:b:====:b:=:b:====:b|}

\multicolumn{11}{l}{}\\

\hhline{~|t:=====:t:=====:t|}
    \multicolumn{1}{l||}{Goal Dependent}
  & \multicolumn{5}{c||}{without Struct Info}
  & \multicolumn{5}{c||}{with Struct Info} \\
\hhline{|t:=::=:t:====::=:t:====:|}
 Prec. class
 & \multicolumn{1}{c||}{O}
 & \multicolumn{1}{c|}{I}
 & \multicolumn{1}{c|}{G}
 & \multicolumn{1}{c|}{F}
 & \multicolumn{1}{c||}{L}
 & \multicolumn{1}{c||}{O}
 & \multicolumn{1}{c|}{I}
 & \multicolumn{1}{c|}{G}
 & \multicolumn{1}{c|}{F}
 & \multicolumn{1}{c||}{L} \\
\hhline{|:=::=::====::=::====:|}
same precision & 96.4 & 96.4 & 96.4 & 96.4 & 96.4 & 90.5 & 90.5 & 90.5 & 90.5 & 90.5 \\
\hhline{||-||-||-|-|-|-||-||-|-|-|-||}
unknown & 3.6 & 3.6 & 3.6 & 3.6 & 3.6 & 9.5 & 9.5 & 9.5 & 9.5 & 9.5 \\
\hhline{|b:=:b:=:b:====:b:=:b:====:b|}
\end{tabular}

\bigskip\bigskip

\begin{tabular}{||l||r|r||r|r||}
\hhline{|t:=:t:==:t:==:t|}
Time diff. class &
\multicolumn{2}{c||}{Goal Ind.} &
\multicolumn{2}{c||}{Goal Dep.} \\
\hhline{|b:=::==::==:|}
\multicolumn{1}{c||}{}
 & \multicolumn{1}{c|}{w/o SI}
 & \multicolumn{1}{c||}{with SI}
 & \multicolumn{1}{c|}{w/o SI}
 & \multicolumn{1}{c||}{with SI} \\
\hhline{|t:=::==::==:|}
$\phantom{0.5 < \mathord{}} \text{degradation} > 1$ & 4.5 & 3.0 & 7.2 & 4.1 \\
 \hhline{||-||-|-||-|-||}
$0.5 < \text{degradation} \leq 1$ & 0.6 & 0.3 & --- & --- \\
 \hhline{||-||-|-||-|-||}
$0.2 < \text{degradation} \leq 0.5$ & 2.4 & 0.9 & 0.5 & 0.5 \\
 \hhline{||-||-|-||-|-||}
$0.1 < \text{degradation} \leq 0.2$ & 1.5 & 0.6 & 0.5 & 0.5 \\
 \hhline{|:=::=:=::=:=:|}
$\phantom{0.5 < \mathord{}} \text{both timed out}$ & 3.0 & 6.3 & 3.6 & 9.5 \\
 \hhline{||-||-|-||-|-||}
$\phantom{0.5 < \mathord{}} \text{same time}$ & 80.7 & 80.7 & 85.5 & 76.9 \\
 \hhline{|:=::=:=::=:=:|}
$0.1 < \text{improvement} \leq 0.2$ & 1.5 & 1.2 & 0.5 & 0.5 \\
 \hhline{||-||-|-||-|-||}
$0.2 < \text{improvement} \leq 0.5$ & 1.8 & 1.2 & 1.4 & 2.3 \\
 \hhline{||-||-|-||-|-||}
$0.5 < \text{improvement} \leq 1$ & 0.9 & 0.6 & --- & 0.9 \\
 \hhline{||-||-|-||-|-||}
$\phantom{0.5 < \mathord{}} \text{improvement} > 1$ & 3.0 & 5.1 & 0.9 & 5.0 \\
\hhline{|b:=:b:==:b:==:b|}
\end{tabular}

\bigskip\bigskip

\begin{tabular}{||c||r|r|r||r|r|r||r|r|r||r|r|r||}
\hhline{|t:=:t:======:t:======:t|}
\multicolumn{1}{||c||}{Total time class}
 & \multicolumn{6}{c||}{Goal Independent}
 & \multicolumn{6}{c||}{Goal Dependent} \\
\hhline{|b:=::===:t:===::===:t:===:|}
\multicolumn{1}{c||}{}
 & \multicolumn{3}{c||}{without SI}
 & \multicolumn{3}{c||}{with SI}
 & \multicolumn{3}{c||}{without SI}
 & \multicolumn{3}{c||}{with SI} \\
\hhline{~|:===::===::===::===:|}
\multicolumn{1}{c||}{}
 & \multicolumn{1}{c|}{\%1} & \multicolumn{1}{c|}{\%2}
 & \multicolumn{1}{c||}{$\Delta$} & \multicolumn{1}{c|}{\%1}
 & \multicolumn{1}{c|}{\%2} & \multicolumn{1}{c||}{$\Delta$}
 & \multicolumn{1}{c|}{\%1} & \multicolumn{1}{c|}{\%2}
 & \multicolumn{1}{c||}{$\Delta$} & \multicolumn{1}{c|}{\%1}
 & \multicolumn{1}{c|}{\%2} & \multicolumn{1}{c||}{$\Delta$} \\
\hhline{|t:=::===::===::===::===:|}
$\text{timed out}$ & 3.3 & 3.3 & --- & 6.6 & 6.6 & --- & 3.6 & 3.6 & --- & 9.5 & 9.5 & --- \\
 \hhline{||-||-|-|-||-|-|-||-|-|-||-|-|-||}
$\phantom{0.5 < \mathord{}} t > 10\phantom{.}$ & 9.0 & 8.1 & -0.9 & 8.4 & 9.0 & 0.6 & 7.2 & 7.7 & 0.5 & 8.6 & 8.1 & -0.5 \\
 \hhline{||-||-|-|-||-|-|-||-|-|-||-|-|-||}
$\phantom{0.}5 < t \leq 10\phantom{.}$ & 0.3 & 0.9 & 0.6 & 1.5 & 1.2 & -0.3 & 1.4 & 0.9 & -0.5 & 1.8 & 2.7 & 0.9 \\
 \hhline{||-||-|-|-||-|-|-||-|-|-||-|-|-||}
$\phantom{0.}1 < t \leq 5\phantom{.5}$ & 7.5 & 7.5 & --- & 6.6 & 6.3 & -0.3 & 3.6 & 3.2 & -0.5 & 5.0 & 4.1 & -0.9 \\
 \hhline{||-||-|-|-||-|-|-||-|-|-||-|-|-||}
$0.5 < t \leq 1\phantom{.5}$ & 2.7 & 2.4 & -0.3 & 3.3 & 3.0 & -0.3 & 5.4 & 5.9 & 0.5 & 3.2 & 3.6 & 0.5 \\
 \hhline{||-||-|-|-||-|-|-||-|-|-||-|-|-||}
$0.2 < t \leq 0.5$ & 8.4 & 9.3 & 0.9 & 10.2 & 10.5 & 0.3 & 13.1 & 12.7 & -0.5 & 13.6 & 13.1 & -0.5 \\
 \hhline{||-||-|-|-||-|-|-||-|-|-||-|-|-||}
$\phantom{0.5 < \mathord{}} t \leq 0.2$ & 68.7 & 68.4 & -0.3 & 63.3 & 63.3 & --- & 65.6 & 66.1 & 0.5 & 58.4 & 58.8 & 0.5 \\
\hhline{|b:=:b:===:b:===:b:===:b:===:b|}
\end{tabular}

\caption{The heuristic based on the number of star-unions.}
\label{tab:starunion-ordering}
\end{table*}

\begin{table*}
\centering
\begin{tabular}{||c||r||r|r|r|r||r||r|r|r|r||}
\hhline{~|t:=====:t:=====:t|}
    \multicolumn{1}{l||}{Goal Independent}
  & \multicolumn{5}{c||}{without Struct Info}
  & \multicolumn{5}{c||}{with Struct Info} \\
\hhline{|t:=::=:t:====::=:t:====:|}
 Prec. class
 & \multicolumn{1}{c||}{O}
 & \multicolumn{1}{c|}{I}
 & \multicolumn{1}{c|}{G}
 & \multicolumn{1}{c|}{F}
 & \multicolumn{1}{c||}{L}
 & \multicolumn{1}{c||}{O}
 & \multicolumn{1}{c|}{I}
 & \multicolumn{1}{c|}{G}
 & \multicolumn{1}{c|}{F}
 & \multicolumn{1}{c||}{L} \\
\hhline{|:=::=::====::=::====:|}
$\phantom{-1}5 < p \leq 10\phantom{-}$ & 0.3 & --- & --- & --- & 0.3 & 0.3 & --- & --- & --- & 0.3 \\
\hhline{||-||-||-|-|-|-||-||-|-|-|-||}
$\phantom{-1}0 < p \leq 2\phantom{-0}$ & 0.9 & --- & --- & --- & 0.9 & 2.7 & 2.4 & --- & --- & 0.3 \\
\hhline{|:=::=::=:=:=:=::=::=:=:=:=:|}
same precision & 94.3 & 95.5 & 96.4 & 96.4 & 95.2 & 89.5 & 90.1 & 93.4 & 93.4 & 92.8 \\
\hhline{||-||-||-|-|-|-||-||-|-|-|-||}
unknown & 3.6 & 3.6 & 3.6 & 3.6 & 3.6 & 6.6 & 6.6 & 6.6 & 6.6 & 6.6 \\
\hhline{|:=::=::=:=:=:=::=::=:=:=:=:|}
$\phantom{1}-2 \leq p < 0\phantom{-2}$ & 0.6 & 0.6 & --- & --- & --- & 0.9 & 0.9 & --- & --- & --- \\
\hhline{||-||-||-|-|-|-||-||-|-|-|-||}
$\phantom{-10 < \mathord{}} p < -20$ & 0.3 & 0.3 & --- & --- & --- & --- & --- & --- & --- & --- \\
\hhline{|b:=:b:=:b:====:b:=:b:====:b|}

\multicolumn{11}{l}{}\\

\hhline{~|t:=====:t:=====:t|}
    \multicolumn{1}{l||}{Goal Dependent}
  & \multicolumn{5}{c||}{without Struct Info}
  & \multicolumn{5}{c||}{with Struct Info} \\
\hhline{|t:=::=:t:====::=:t:====:|}
 Prec. class
 & \multicolumn{1}{c||}{O}
 & \multicolumn{1}{c|}{I}
 & \multicolumn{1}{c|}{G}
 & \multicolumn{1}{c|}{F}
 & \multicolumn{1}{c||}{L}
 & \multicolumn{1}{c||}{O}
 & \multicolumn{1}{c|}{I}
 & \multicolumn{1}{c|}{G}
 & \multicolumn{1}{c|}{F}
 & \multicolumn{1}{c||}{L} \\
\hhline{|:=::=::====::=::====:|}
$\phantom{-1}0 < p \leq 2\phantom{-0}$ & 0.5 & --- & --- & --- & 0.5 & --- & --- & --- & --- & --- \\
\hhline{|:=::=::=:=:=:=::=::=:=:=:=:|}
same precision & 94.6 & 95.0 & 95.5 & 95.5 & 95.0 & 89.6 & 89.6 & 89.6 & 89.6 & 89.6 \\
\hhline{||-||-||-|-|-|-||-||-|-|-|-||}
unknown & 4.5 & 4.5 & 4.5 & 4.5 & 4.5 & 10.4 & 10.4 & 10.4 & 10.4 & 10.4 \\
\hhline{|:=::=::=:=:=:=::=::=:=:=:=:|}
$-20 \leq p < -10$ & 0.5 & 0.5 & --- & --- & --- & --- & --- & --- & --- & --- \\
\hhline{|b:=:b:=:b:====:b:=:b:====:b|}
\end{tabular}

\bigskip\bigskip

\begin{tabular}{||l||r|r||r|r||}
\hhline{|t:=:t:==:t:==:t|}
Time diff. class &
\multicolumn{2}{c||}{Goal Ind.} &
\multicolumn{2}{c||}{Goal Dep.} \\
\hhline{|b:=::==::==:|}
\multicolumn{1}{c||}{} &
\multicolumn{1}{c|}{w/o SI} &
\multicolumn{1}{c||}{with SI} &
\multicolumn{1}{c|}{w/o SI} &
\multicolumn{1}{c||}{with SI} \\
\hhline{|t:=::==::==:|}
$\phantom{0.5 < \mathord{}} \text{degradation} > 1$ & 6.9 & 4.8 & 8.1 & 7.7 \\
 \hhline{||-||-|-||-|-||}
$0.5 < \text{degradation} \leq 1$ & 2.1 & 1.5 & 1.8 & 0.5 \\
 \hhline{||-||-|-||-|-||}
$0.2 < \text{degradation} \leq 0.5$ & 2.4 & 1.8 & 1.8 & 2.7 \\
 \hhline{||-||-|-||-|-||}
$0.1 < \text{degradation} \leq 0.2$ & 1.2 & 3.3 & 2.3 & 3.2 \\
 \hhline{|:=::=:=::=:=:|}
$\phantom{0.5 < \mathord{}} \text{both timed out}$ & 2.4 & 5.7 & 3.6 & 9.0 \\
 \hhline{||-||-|-||-|-||}
$\phantom{0.5 < \mathord{}} \text{same time}$ & 77.4 & 73.5 & 78.7 & 71.9 \\
 \hhline{|:=::=:=::=:=:|}
$0.1 < \text{improvement} \leq 0.2$ & 1.2 & 0.3 & --- & --- \\
 \hhline{||-||-|-||-|-||}
$0.2 < \text{improvement} \leq 0.5$ & 0.6 & 1.8 & 0.9 & 0.9 \\
 \hhline{||-||-|-||-|-||}
$0.5 < \text{improvement} \leq 1$ & 0.9 & --- & 0.5 & --- \\
 \hhline{||-||-|-||-|-||}
$\phantom{0.5 < \mathord{}} \text{improvement} > 1$ & 4.8 & 7.2 & 2.3 & 4.1 \\
\hhline{|b:=:b:==:b:==:b|}
\end{tabular}

\bigskip\bigskip

\begin{tabular}{||c||r|r|r||r|r|r||r|r|r||r|r|r||}
\hhline{|t:=:t:======:t:======:t|}
\multicolumn{1}{||c||}{Total time class}
 & \multicolumn{6}{c||}{Goal Independent}
 & \multicolumn{6}{c||}{Goal Dependent} \\
\hhline{|b:=::===:t:===::===:t:===:|}
\multicolumn{1}{c||}{}
 & \multicolumn{3}{c||}{without SI}
 & \multicolumn{3}{c||}{with SI}
 & \multicolumn{3}{c||}{without SI}
 & \multicolumn{3}{c||}{with SI} \\
\hhline{~|:===::===::===::===:|}
\multicolumn{1}{c||}{}
 & \multicolumn{1}{c|}{\%1} & \multicolumn{1}{c|}{\%2}
 & \multicolumn{1}{c||}{$\Delta$} & \multicolumn{1}{c|}{\%1}
 & \multicolumn{1}{c|}{\%2} & \multicolumn{1}{c||}{$\Delta$}
 & \multicolumn{1}{c|}{\%1} & \multicolumn{1}{c|}{\%2}
 & \multicolumn{1}{c||}{$\Delta$} & \multicolumn{1}{c|}{\%1}
 & \multicolumn{1}{c|}{\%2} & \multicolumn{1}{c||}{$\Delta$} \\
\hhline{|t:=::===::===::===::===:|}
$\text{timed out}$ & 3.3 & 2.7 & -0.6 & 6.6 & 5.7 & -0.9 & 3.6 & 4.5 & 0.9 & 9.5 & 10.0 & 0.5 \\
 \hhline{||-||-|-|-||-|-|-||-|-|-||-|-|-||}
$\phantom{0.5 < \mathord{}} t > 10\phantom{.}$ & 9.0 & 9.6 & 0.6 & 8.4 & 8.7 & 0.3 & 7.2 & 6.8 & -0.5 & 8.6 & 7.7 & -0.9 \\
 \hhline{||-||-|-|-||-|-|-||-|-|-||-|-|-||}
$\phantom{0.}5 < t \leq 10\phantom{.}$ & 0.3 & 2.1 & 1.8 & 1.5 & 1.8 & 0.3 & 1.4 & 1.4 & --- & 1.8 & 2.7 & 0.9 \\
 \hhline{||-||-|-|-||-|-|-||-|-|-||-|-|-||}
$\phantom{0.}1 < t \leq 5\phantom{.5}$ & 7.5 & 6.0 & -1.5 & 6.6 & 6.9 & 0.3 & 3.6 & 4.5 & 0.9 & 5.0 & 5.0 & --- \\
 \hhline{||-||-|-|-||-|-|-||-|-|-||-|-|-||}
$0.5 < t \leq 1\phantom{.5}$ & 2.7 & 3.0 & 0.3 & 3.3 & 3.9 & 0.6 & 5.4 & 4.1 & -1.4 & 3.2 & 3.6 & 0.5 \\
 \hhline{||-||-|-|-||-|-|-||-|-|-||-|-|-||}
$0.2 < t \leq 0.5$ & 8.4 & 9.9 & 1.5 & 10.2 & 13.3 & 3.0 & 13.1 & 13.1 & --- & 13.6 & 15.4 & 1.8 \\
 \hhline{||-||-|-|-||-|-|-||-|-|-||-|-|-||}
$\phantom{0.5 < \mathord{}} t \leq 0.2$ & 68.7 & 66.6 & -2.1 & 63.3 & 59.6 & -3.6 & 65.6 & 65.6 & --- & 58.4 & 55.7 & -2.7 \\
\hhline{|b:=:b:===:b:===:b:===:b:===:b|}
\end{tabular}

\caption{The heuristic based on freeness and linearity.}
\label{tab:maxfreelin-ordering}
\end{table*}

In order to evaluate the effects of these two heuristics
on real programs, we have implemented and compared them
with respect to the ``straight'' abstract computation,
which considers the non-grounding bindings using the
left-to-right order.\footnote{The base domain
is $\Pos \times \SFL$, both with and without structural information.}
The results reported
in Tables~\ref{tab:starunion-ordering} and~\ref{tab:maxfreelin-ordering}
can be summarized as follows:
\begin{enumerate}
\item
the precision on the groundness and freeness components
is not affected;
\item
the precision on the independent pairs and linearity components
is rarely affected,
in particular when considering goal-dependent analyses;
\item
even for real programs, as was the case for the artificial examples
given above, the precision can be increased as well as decreased.
\end{enumerate}

Looking at
Tables~\ref{tab:starunion-ordering} and~\ref{tab:maxfreelin-ordering},
it can be seen that the heuristic based on freeness and linearity information
is slightly better than the use of the straight order,
which, in its turn, is slightly better than the heuristic
based on the number of star-unions.

Clearly, since these results could not be generalized to other
orderings, our investigation cannot be considered really conclusive.
Besides designing ``smarter'' heuristics,
it would be interesting to provide a kind of
\emph{responsiveness test} for the underlying domain
with respect to the choice of ordering for the non-grounding bindings:
a simple test consists in measuring how much the precision can be affected,
in either way, by the application of an almost arbitrary order.
This is the motivation for the comparison reported in
Table~\ref{tab:reverse-ordering},
where the order is from right-to-left, the reverse of the usual one.
As for the results given in
Tables~\ref{tab:starunion-ordering} and~\ref{tab:maxfreelin-ordering},
the number of changes to the precision observed
in Table~\ref{tab:reverse-ordering} is small
and all the observations made above still hold.
Surprisingly, this reversed ordering
provides marginally better precision results than those
obtained using the considered heuristics.\footnote{%
It is worth noting that the only precision improvement reported
in Table~\ref{tab:reverse-ordering} for the goal-dependent analysis
with structural information (caused by the program \texttt{semi})
corresponds to the precision decrease reported
in Table~\ref{tab:nostruct-vs-struct}.
This confirms that, as informally discussed in Section~\ref{sec:struct},
such a precision decrease was due to the non-commutativity
of the $\amgu$ operator on $\Pos \times \SFL$.}

\begin{table*}
\centering
\begin{tabular}{||c||r||r|r|r|r||r||r|r|r|r||}
\hhline{~|t:=====:t:=====:t|}
    \multicolumn{1}{l||}{Goal Independent}
  & \multicolumn{5}{c||}{without Struct Info}
  & \multicolumn{5}{c||}{with Struct Info} \\
\hhline{|t:=::=:t:====::=:t:====:|}
 Prec. class
 & \multicolumn{1}{c||}{O}
 & \multicolumn{1}{c|}{I}
 & \multicolumn{1}{c|}{G}
 & \multicolumn{1}{c|}{F}
 & \multicolumn{1}{c||}{L}
 & \multicolumn{1}{c||}{O}
 & \multicolumn{1}{c|}{I}
 & \multicolumn{1}{c|}{G}
 & \multicolumn{1}{c|}{F}
 & \multicolumn{1}{c||}{L} \\
\hhline{|:=::=::====::=::====:|}
$\phantom{-1}5 < p \leq 10\phantom{-}$ & 0.3 & --- & --- & --- & 0.3 & 0.3 & --- & --- & --- & 0.3 \\
\hhline{||-||-||-|-|-|-||-||-|-|-|-||}
$\phantom{-1}0 < p \leq 2\phantom{-0}$ & 0.9 & 0.3 & --- & --- & 0.6 & 4.2 & 3.0 & --- & --- & 1.2 \\
\hhline{|:=::=::=:=:=:=::=::=:=:=:=:|}
same precision & 94.3 & 95.2 & 96.4 & 96.4 & 95.5 & 87.7 & 89.2 & 93.4 & 93.4 & 91.9 \\
\hhline{||-||-||-|-|-|-||-||-|-|-|-||}
unknown & 3.6 & 3.6 & 3.6 & 3.6 & 3.6 & 6.6 & 6.6 & 6.6 & 6.6 & 6.6 \\
\hhline{|:=::=::=:=:=:=::=::=:=:=:=:|}
$\phantom{1}-2 \leq p < 0\phantom{-2}$ & 0.6 & 0.6 & --- & --- & --- & 1.2 & 1.2 & --- & --- & --- \\
\hhline{||-||-||-|-|-|-||-||-|-|-|-||}
$\phantom{-10 < \mathord{}} p < -20$ & 0.3 & 0.3 & --- & --- & --- & --- & --- & --- & --- & --- \\
\hhline{|b:=:b:=:b:====:b:=:b:====:b|}

\multicolumn{11}{l}{}\\

\hhline{~|t:=====:t:=====:t|}
    \multicolumn{1}{l||}{Goal Dependent}
  & \multicolumn{5}{c||}{without Struct Info}
  & \multicolumn{5}{c||}{with Struct Info} \\
\hhline{|t:=::=:t:====::=:t:====:|}
 Prec. class
 & \multicolumn{1}{c||}{O}
 & \multicolumn{1}{c|}{I}
 & \multicolumn{1}{c|}{G}
 & \multicolumn{1}{c|}{F}
 & \multicolumn{1}{c||}{L}
 & \multicolumn{1}{c||}{O}
 & \multicolumn{1}{c|}{I}
 & \multicolumn{1}{c|}{G}
 & \multicolumn{1}{c|}{F}
 & \multicolumn{1}{c||}{L} \\
\hhline{|:=::=::====::=::====:|}
$\phantom{-1}0 < p \leq 2\phantom{-0}$ & 0.5 & --- & --- & --- & 0.5 & 0.5 & --- & --- & --- & 0.5 \\
\hhline{|:=::=::=:=:=:=::=::=:=:=:=:|}
same precision & 95.5 & 95.9 & 96.4 & 96.4 & 95.9 & 90.0 & 90.5 & 90.5 & 90.5 & 90.0 \\
\hhline{||-||-||-|-|-|-||-||-|-|-|-||}
unknown & 3.6 & 3.6 & 3.6 & 3.6 & 3.6 & 9.5 & 9.5 & 9.5 & 9.5 & 9.5 \\
\hhline{|:=::=::=:=:=:=::=::=:=:=:=:|}
$-20 \leq p < -10$ & 0.5 & 0.5 & --- & --- & --- & --- & --- & --- & --- & --- \\
\hhline{|b:=:b:=:b:====:b:=:b:====:b|}
\end{tabular}

\bigskip\bigskip

\begin{tabular}{||l||r|r||r|r||}
\hhline{|t:=:t:==:t:==:t|}
Time diff. class &
\multicolumn{2}{c||}{Goal Ind.} &
\multicolumn{2}{c||}{Goal Dep.} \\
\hhline{|b:=::==::==:|}
\multicolumn{1}{c||}{} &
\multicolumn{1}{c|}{w/o SI} &
\multicolumn{1}{c||}{with SI} &
\multicolumn{1}{c|}{w/o SI} &
\multicolumn{1}{c||}{with SI} \\
\hhline{|t:=::==::==:|}
$\phantom{0.5 < \mathord{}} \text{degradation} > 1$ & 4.2 & 6.0 & 4.5 & 6.8 \\
 \hhline{||-||-|-||-|-||}
$0.5 < \text{degradation} \leq 1$ & 0.6 & 0.6 & --- & --- \\
 \hhline{||-||-|-||-|-||}
$0.2 < \text{degradation} \leq 0.5$ & 2.4 & 1.5 & 1.4 & 0.9 \\
 \hhline{||-||-|-||-|-||}
$0.1 < \text{degradation} \leq 0.2$ & 1.8 & 0.9 & 0.5 & --- \\
 \hhline{|:=::=:=::=:=:|}
$\phantom{0.5 < \mathord{}} \text{both timed out}$ & 2.4 & 5.7 & 3.6 & 9.0 \\
 \hhline{||-||-|-||-|-||}
$\phantom{0.5 < \mathord{}} \text{same time}$ & 78.3 & 76.2 & 82.8 & 74.2 \\
 \hhline{|:=::=:=::=:=:|}
$0.1 < \text{improvement} \leq 0.2$ & 1.5 & 1.2 & 1.8 & 0.9 \\
 \hhline{||-||-|-||-|-||}
$0.2 < \text{improvement} \leq 0.5$ & 1.8 & 0.3 & 1.4 & 1.8 \\
 \hhline{||-||-|-||-|-||}
$0.5 < \text{improvement} \leq 1$ & 0.9 & 0.9 & 0.5 & 0.5 \\
 \hhline{||-||-|-||-|-||}
$\phantom{0.5 < \mathord{}} \text{improvement} > 1$ & 6.0 & 6.6 & 3.6 & 5.9 \\
\hhline{|b:=:b:==:b:==:b|}
\end{tabular}

\bigskip\bigskip

\begin{tabular}{||c||r|r|r||r|r|r||r|r|r||r|r|r||}
\hhline{|t:=:t:======:t:======:t|}
\multicolumn{1}{||c||}{Total time class}
 & \multicolumn{6}{c||}{Goal Independent}
 & \multicolumn{6}{c||}{Goal Dependent} \\
\hhline{|b:=::===:t:===::===:t:===:|}
\multicolumn{1}{c||}{}
 & \multicolumn{3}{c||}{without SI}
 & \multicolumn{3}{c||}{with SI}
 & \multicolumn{3}{c||}{without SI}
 & \multicolumn{3}{c||}{with SI} \\
\hhline{~|:===::===::===::===:|}
\multicolumn{1}{c||}{}
 & \multicolumn{1}{c|}{\%1} & \multicolumn{1}{c|}{\%2}
 & \multicolumn{1}{c||}{$\Delta$} & \multicolumn{1}{c|}{\%1}
 & \multicolumn{1}{c|}{\%2} & \multicolumn{1}{c||}{$\Delta$}
 & \multicolumn{1}{c|}{\%1} & \multicolumn{1}{c|}{\%2}
 & \multicolumn{1}{c||}{$\Delta$} & \multicolumn{1}{c|}{\%1}
 & \multicolumn{1}{c|}{\%2} & \multicolumn{1}{c||}{$\Delta$} \\
\hhline{|t:=::===::===::===::===:|}
$\text{timed out}$ & 3.3 & 2.7 & -0.6 & 6.6 & 5.7 & -0.9 & 3.6 & 3.6 & --- & 9.5 & 9.0 & -0.5 \\
 \hhline{||-||-|-|-||-|-|-||-|-|-||-|-|-||}
$\phantom{0.5 < \mathord{}} t > 10\phantom{.}$ & 9.0 & 8.7 & -0.3 & 8.4 & 9.9 & 1.5 & 7.2 & 7.7 & 0.5 & 8.6 & 8.1 & -0.5 \\
 \hhline{||-||-|-|-||-|-|-||-|-|-||-|-|-||}
$\phantom{0.}5 < t \leq 10\phantom{.}$ & 0.3 & 1.8 & 1.5 & 1.5 & 1.5 & --- & 1.4 & 0.5 & -0.9 & 1.8 & 2.7 & 0.9 \\
 \hhline{||-||-|-|-||-|-|-||-|-|-||-|-|-||}
$\phantom{0.}1 < t \leq 5\phantom{.5}$ & 7.5 & 6.9 & -0.6 & 6.6 & 6.0 & -0.6 & 3.6 & 3.2 & -0.5 & 5.0 & 4.5 & -0.5 \\
 \hhline{||-||-|-|-||-|-|-||-|-|-||-|-|-||}
$0.5 < t \leq 1\phantom{.5}$ & 2.7 & 2.4 & -0.3 & 3.3 & 2.7 & -0.6 & 5.4 & 5.4 & --- & 3.2 & 3.6 & 0.5 \\
 \hhline{||-||-|-|-||-|-|-||-|-|-||-|-|-||}
$0.2 < t \leq 0.5$ & 8.4 & 8.7 & 0.3 & 10.2 & 11.1 & 0.9 & 13.1 & 13.1 & --- & 13.6 & 12.2 & -1.4 \\
 \hhline{||-||-|-|-||-|-|-||-|-|-||-|-|-||}
$\phantom{0.5 < \mathord{}} t \leq 0.2$ & 68.7 & 68.7 & --- & 63.3 & 63.0 & -0.3 & 65.6 & 66.5 & 0.9 & 58.4 & 59.7 & 1.4 \\
\hhline{|b:=:b:===:b:===:b:===:b:===:b|}
\end{tabular}

\caption{Reversing the ordering of the non-grounding bindings.}
\label{tab:reverse-ordering}
\end{table*}

\section{The Reduced Product between Pos and Sharing}
\label{sec:enhancedpos}

The overlap between the information provided by $\Pos$
and the information provided by $\Sharing$ mentioned in
Section~\ref{sec:simplepos} means that
the Cartesian product $\Pos \times \SFL$ contains redundancy,
that is, there is more than one element that can characterize the same set
of concrete computational states.

In \cite{BagnaraZH00}, two techniques that are able to remove
some of this redundancy were experimentally evaluated.
One of these aims at identifying those pairs of variables $(x,y)$
for which the Boolean formula of the $\Pos$ component implies
the \emph{binary disjunction} $x \vee y$.
In such a case, it is always safe to assume that
the variables $x$ and $y$ are independent.\footnote{Note that
this observation dates back, at least, to \cite{CrnogoracKS96}.}
Since the number of independent pairs is one of the quantities
explicitly measured,
this enhancement has the potential for ``immediate'' precision gains.
The other technique exploits the knowledge of the sets
of \emph{ground-equivalent} variables:
the variables in $e \sseq \VI$ are ground-equivalent in $\phi \in \Pos$
if and only if, for each $x,y \in e$, $\phi \models (x \piff y)$.
For a description of how
these sets can be used to improve sharing analysis,
the reader is referred to \cite{BagnaraZH00}.
The main motivation for experimenting with this specific reduction was
the ease of its implementation, since all the
needed information can easily be recovered from the already
computed $E$ component of the GER implementation of $\Pos$
\cite{BagnaraS99}.
The experimental evaluation results given in \cite{BagnaraZH00} for
these two techniques show precision improvements
with only three of the programs and, also, only with respect to
the number of independent pairs that were found.
Those results just apply to these limited forms of reduction,
so could not be considered a complete account of all the
possible precision gains.

\begin{table*}
\centering
\begin{tabular}{||c||r||r|r|r|r||r||r|r|r|r||}
\hhline{~|t:=====:t:=====:t|}
    \multicolumn{1}{l||}{Goal Independent}
  & \multicolumn{5}{c||}{without Struct Info}
  & \multicolumn{5}{c||}{with Struct Info} \\
\hhline{|t:=::=:t:====::=:t:====:|}
 Prec. class
 & \multicolumn{1}{c||}{O}
 & \multicolumn{1}{c|}{I}
 & \multicolumn{1}{c|}{G}
 & \multicolumn{1}{c|}{F}
 & \multicolumn{1}{c||}{L}
 & \multicolumn{1}{c||}{O}
 & \multicolumn{1}{c|}{I}
 & \multicolumn{1}{c|}{G}
 & \multicolumn{1}{c|}{F}
 & \multicolumn{1}{c||}{L} \\
\hhline{|:=::=::====::=::====:|}
$\phantom{-1}5 < p \leq 10\phantom{-}$ & --- & --- & --- & --- & --- & 0.3 & 0.3 & --- & --- & --- \\
\hhline{||-||-||-|-|-|-||-||-|-|-|-||}
$\phantom{-1}2 < p \leq 5\phantom{-0}$ & 0.3 & 0.3 & --- & --- & --- & --- & --- & --- & --- & --- \\
\hhline{||-||-||-|-|-|-||-||-|-|-|-||}
$\phantom{-1}0 < p \leq 2\phantom{-0}$ & 2.7 & 2.7 & --- & --- & 0.6 & 3.9 & 3.9 & --- & --- & 0.6 \\
\hhline{|:=::=::=:=:=:=::=::=:=:=:=:|}
same precision & 86.1 & 86.1 & 89.2 & 89.2 & 88.6 & 80.7 & 80.7 & 84.9 & 84.9 & 84.3 \\
\hhline{||-||-||-|-|-|-||-||-|-|-|-||}
unknown & 10.8 & 10.8 & 10.8 & 10.8 & 10.8 & 15.1 & 15.1 & 15.1 & 15.1 & 15.1 \\
\hhline{|b:=:b:=:b:====:b:=:b:====:b|}

\multicolumn{11}{l}{}\\

\hhline{~|t:=====:t:=====:t|}
    \multicolumn{1}{l||}{Goal Dependent}
  & \multicolumn{5}{c||}{without Struct Info}
  & \multicolumn{5}{c||}{with Struct Info} \\
\hhline{|t:=::=:t:====::=:t:====:|}
 Prec. class
 & \multicolumn{1}{c||}{O}
 & \multicolumn{1}{c|}{I}
 & \multicolumn{1}{c|}{G}
 & \multicolumn{1}{c|}{F}
 & \multicolumn{1}{c||}{L}
 & \multicolumn{1}{c||}{O}
 & \multicolumn{1}{c|}{I}
 & \multicolumn{1}{c|}{G}
 & \multicolumn{1}{c|}{F}
 & \multicolumn{1}{c||}{L} \\
\hhline{|:=::=::====::=::====:|}
$\phantom{-20 < \mathord{}} p > 20\phantom{-}$ & 0.5 & 0.5 & --- & --- & --- & --- & --- & --- & --- & --- \\
\hhline{||-||-||-|-|-|-||-||-|-|-|-||}
$\phantom{-}10 < p \leq 20\phantom{-}$ & --- & --- & --- & --- & --- & 0.5 & 0.5 & --- & --- & --- \\
\hhline{||-||-||-|-|-|-||-||-|-|-|-||}
$\phantom{-1}5 < p \leq 10\phantom{-}$ & --- & --- & --- & --- & --- & 0.5 & 0.5 & --- & --- & --- \\
\hhline{||-||-||-|-|-|-||-||-|-|-|-||}
$\phantom{-1}0 < p \leq 2\phantom{-0}$ & 2.7 & 2.7 & --- & --- & --- & 2.7 & 2.7 & --- & --- & --- \\
\hhline{|:=::=::=:=:=:=::=::=:=:=:=:|}
same precision & 89.1 & 89.1 & 92.3 & 92.3 & 92.3 & 77.8 & 77.8 & 81.4 & 81.4 & 81.4 \\
\hhline{||-||-||-|-|-|-||-||-|-|-|-||}
unknown & 7.7 & 7.7 & 7.7 & 7.7 & 7.7 & 18.6 & 18.6 & 18.6 & 18.6 & 18.6 \\
\hhline{|b:=:b:=:b:====:b:=:b:====:b|}
\end{tabular}

\caption{$\Pos \times \PSDFL$ versus $\Pos \otimes \SFL$.}
\label{tab:simplepos-vs-enhancedpos-sfl}
\end{table*}

The full reduced product \cite{CousotC79} between $\Pos$ and $\Sharing$
has been elegantly characterized in \cite{CodishSS99},
where set-sharing \emph{\`a la} Jacobs and Langen is expressed
in terms of elements of the $\Pos$ domain itself.
Let $\Models{\phi}{\VI}$ denote the set of all the models of
the Boolean function $\phi$ defined over the set of variables $\VI$.
Then, the isomorphism maps each set-sharing element $\sh \in \SH$
into the Boolean formula $\phi \in \Pos$ such that
\[
  \Models{\phi}{\VI}
    = \{\, \VI \setdiff S \mid S \in \sh \,\}
      \union \{ \VI \}.
\]
The sharing information encoded by an element
$(\phi_g, \phi_{\sh}) \in \Pos \times \Pos$
can be improved by replacing the second component
(that is, the Boolean formula describing set-sharing information)
with the conjunction $\phi_g \wedge \phi_{\sh}$.
The reader is referred to \cite{CodishSS99} for a complete
account of this composition and a justification of its correctness.

This specification of the reduced product can be reformulated,
using the standard set-sharing
representation for the second component, to define
a reduction procedure
$\fund{\reduce}{\Pos \times \SH}{\SH}$ such that,
for all $\phi_g \in \Pos$, $\sh \in \SH$,
\[
  \reduce(\phi_g, \sh)
    = \bigl\{\,
	S \in \sh
      \bigm|
	(\VI \setdiff S) \in \Models{\phi_g}{\VI}
      \,\bigr\}.
\]

The enhanced integration of $\Pos$ and $\SFL$,
based on the above reduction operator,
is denoted here by $\Pos \otimes \SFL$.
From a formal point of view,
this is \emph{not} the reduced product between $\Pos$ and $\SFL$:
while there is a complete reduction between $\Pos$ and $\SH$,
the same does not necessarily hold for the combination
with freeness and linearity information.
Also note that the domain $\Pos \otimes \SFL$
is strictly more precise than the domain $\ScozzariShPSh$,
defined in~\cite{Scozzari00} for pair-sharing analysis.
This is because the domain $\ScozzariShPSh$ is the reduced product
of a strict abstraction of $\Pos$ and a strict abstraction of $\SH$.

When using the domain $\PSD$ in place of $\SH$,
the `$\reduce$' operator specified above can interact in subtle ways
with an implementation removing the $\rho$-redundant sharing groups
from the elements of $\PSD$.
The following is an example where such an interaction provides
results that are not correct.

Let $\VI = \{x, y, z\}$ and $\sh = \{ xy, xz, yz, xyz \} \in \PSD$
be the current set-sharing description.
Suppose that the implementation internally represents $\sh$
by using the $\rho$-reduced element
$\sh_\reduced = \{ xy, xz, yz \}$, so that $\sh = \rho(\sh_\reduced)$.
Suppose also that the groundness description
computed on the domain $\Pos$ is
$\phi_g = (x \piff y \piff z)$.
Note that we have
\(
  \Models{\phi_g}{\VI} = \bigl\{ \emptyset, \{x,y,z\} \bigr\}
\).
Then we have
\begin{align*}
  \sh'
    &= \reduce(\sh, \phi_g)
     = \{ xyz \}; \\
  \sh'_\reduced
    &= \reduce(\sh_\reduced, \phi_g)
     = \emptyset.
\end{align*}
The two $\Pos$-reduced elements $\sh'$ and $\sh'_\reduced$
are not equivalent, even modulo $\rho$.

Note that the above example does not mean that the reduced product
between $\Pos$ and $\PSD$ yields results that are not correct;
neither does it mean that it is less precise
than the reduced product between $\Pos$ and $\SH$
for the computation of the observables.
More simply, the optimizations used
in our current implementation of $\PSD$
are not compatible with the above reduction process.
Therefore, in Table~\ref{tab:simplepos-vs-enhancedpos-sfl}
we show the precision results obtained
when comparing the base domain $\Pos \times \PSDFL$
with the domain $\Pos \otimes \SFL$:
the implementation of $\Pos \otimes \SFL$,
by avoiding $\rho$-reductions,
is not affected by the correctness problem mentioned above.

The precision comparison provides empirical evidence
that $\Pos \otimes \SFL$ is more effective than
the combination considered in \cite{BagnaraZH00}.
However, as indicated by the number of time-outs reported
in Table~\ref{tab:simplepos-vs-enhancedpos-sfl},
using $\Pos \otimes \SFL$ is not feasible
due to its intrinsic exponential complexity.
We deliberately decided not to include the time comparison, since
it would have provided no information at all:
the efficiency degradations, which are largely caused by the lack
of $\rho$-reductions, should not be attributed to the enhanced
combination with $\Pos$.
In this respect, the reader looking for more details is referred
to~\cite{BagnaraHZ02TCS}.

For the only purpose of investigating how many precision improvements
may have been missed in the previous comparison due to the high
number of time-outs, we have performed another experimental evaluation
where we have compared the base domain $\Pos \times \PSDFL$
and the domain $\Pos \otimes \PSDFL$.
We stress the fact that, given the observation made previously,
such a precision comparison provides an \emph{over-estimation}
for the actual improvements that can be obtained by a correct
integration of the $\rho$-reduction and the `$\reduce$' operators.
A detailed investigation of the experimental data,
which cannot be reported here for space reasons,
has shown that the number of precision improvements shown in
Table~\ref{tab:simplepos-vs-enhancedpos-sfl} could at most double.
In particular, improvements are more likely to occur for
goal-independent analyses.

\section{Ground-or-free Variables}
\label{sec:ground-or-free}

Most of the ideas investigated in the present work are based on earlier
work by other authors.
In this section, we describe one originally proposed
in~\cite{BagnaraZH00}.
Consider the analysis of the binding $x = t$ and suppose that,
on a set of computation paths, this binding is reached with $x$ ground while,
on the remaining computation paths, the binding is reached with $x$ free.
In both cases $x$ will be linear and this is all that will be recorded
when using the usual combination $\Pos \times \SFL$.
This information is valuable since,
in the case that $x$ and $t$ are independent,
it allows the star-union operation for the relevant component
for $t$ to be dispensed with.
However, the information that is lost, that is, $x$ being either
ground or free, is equally valuable, since this would allow
the avoidance of the star-union of \emph{both}
the relevant components for $x$ and $t$,
even when $x$ and $t$ may share.
This loss has the disadvantages that
CPU time is wasted by performing unnecessary but costly operations
and that the precision is potentially degraded:
not only are the extra star-unions useless
for correctness but may introduce redundant
sharing groups to the detriment of accuracy.
It is therefore useful to track the additional mode `ground-or-free'.

The analysis domain $\SFL$ is extended with the component
$\GF \defeq \wp(\VI)$ consisting of the set of variables
that are known to be either ground or free.
As for freeness and linearity, the approximation ordering on $\GF$
is given by reverse subset inclusion.
When computing the abstract $\mgu$ on the new domain
\[
  \SGFL \defeq \SH \times F \times \GF \times L,
\]
the property of being ground-or-free
is used and propagated in almost the same way as freeness information.

\begin{defn} \summary{(Improved abstract operations over $\SGFL$.)}
\label{def:abs-funcs-SGFL}
Let $\sgfl = \langle \sh, f, \gf, l \rangle \in \SGFL$.
We define the predicate $\fund{\gfree_{\sgfl}}{\Terms}{\Bool}$
such that, for each first order term $t$,
where $V_t \defeq \vars(t) \sseq \VI$,
\[
  \gfree_{\sgfl}(t)
    \defeq
      \bigl(\rel(V_t, \sh) = \emptyset \bigr)
        \lor
      (\exists x \in \VI \st x = t \land x \in \gf).
\]
Consider the specification of the abstract operations over $\SFL$
given in Definition~\ref{def:abs-funcs-SFL}. The improved operator
$\fund{\amgu}{\SGFL \times \Bind}{\SGFL}$ is given by
\begin{align*}
  \amgu\bigl(d, x = t\bigr)
    &\defeq
      \langle \sh', f', \gf', l' \rangle,
\end{align*}
where $f'$ and $l''$ are defined as in Definition~\ref{def:abs-funcs-SFL}
and
\begin{align*}
  \sh' &= \irel(V_{xt},\sh) \union \bin\bigl(S_x, S_t\bigr);\\
  S_x  &= \begin{cases}
            R_x,	&\text{if $\gfree_{\sgfl}(x)
                                   \lor
                                   \gfree_{\sgfl}(t)
                                   \lor
                                   \bigl(
                                     \lin_{\sgfl}(t)
                                     \land
                                     \ind_{\sgfl}(x, t)
                                   \bigr)$;} \\
            R_x^\star, &\text{otherwise;}
          \end{cases}\\
  S_t  &= \begin{cases}
            R_t,       &\text{if $\gfree_{\sgfl}(x)
                                  \lor \gfree_{\sgfl}(t)
                                  \lor \bigl(
                                         \lin_{\sgfl}(x)
                                         \land
                                         \ind_{\sgfl}(x, t)
                                       \bigr)$;} \\
            R_t^\star, &\text{otherwise;}
          \end{cases} \\
\gf'   &= \bigl( \VI \setdiff vars(\sh') \bigr) \union \gf''; \\
\gf''  &= \begin{cases}
            \gf,                      &\text{if $\gfree_{\sgfl}(x)
                                                   \land
                                                 \gfree_{\sgfl}(t)$;} \\
            \gf \setdiff \vars(R_x),  &\text{if $\gfree_{\sgfl}(x)$;} \\
            \gf \setdiff \vars(R_t),  &\text{if $\gfree_{\sgfl}(t)$;} \\
            \gf \setdiff \vars(R_x \union R_t),
                                      &\text{otherwise;}
          \end{cases} \\
  l'   &= \gf' \union l''.
\end{align*}
\end{defn}

The computation of the set $\gf''$ is very similar to
the computation of the set $f'$
as given in Definition~\ref{def:abs-funcs-SFL}.
The new ground-or-free component $\gf'$ is obtained by adding to $\gf''$
the set of all the ground variables:
in other words, if a variable ``loses freeness'' then it also loses
its ground-or-free status unless it is known to be definitely ground.
It can be noted that,
in the computation of this improved $\amgu$,
the ground-or-free property takes the role previously played by freeness.
In particular, when computing $\sh'$, all the tests for freeness
have been replaced by tests on the newly defined Boolean function $\gfree_d$;
similarly, in the computation of the new linearity component $l'$,
the set $f'$ has been replaced by $\gf'$ (since any ground-or-free
variable is also linear).
It is also easy to generalize the improvement
for definitely cyclic bindings introduced
in Definition~\ref{def:abs-funcs-SFL-cyclic} to the domain $\SGFL$:
as before, the test $\free_{\sfl}(x)$ needs to be replaced
with the new test $\gfree_{\sgfl}(x)$.

\begin{table*}
\centering

\begin{tabular}{||c||r||r|r|r|r|r||r||r|r|r|r|r||}
\hhline{~|t:======:t:======:t|}
    \multicolumn{1}{l||}{Goal Ind.}
  & \multicolumn{6}{c||}{without Struct Info}
  & \multicolumn{6}{c||}{with Struct Info} \\
\hhline{|t:=::=:t:=====::=:t:=====:|}
 Prec. class
 & \multicolumn{1}{c||}{O}
 & \multicolumn{1}{c|}{I}
 & \multicolumn{1}{c|}{G}
 & \multicolumn{1}{c|}{F}
 & \multicolumn{1}{c|}{GF}
 & \multicolumn{1}{c||}{L}
 & \multicolumn{1}{c||}{O}
 & \multicolumn{1}{c|}{I}
 & \multicolumn{1}{c|}{G}
 & \multicolumn{1}{c|}{F}
 & \multicolumn{1}{c|}{GF}
 & \multicolumn{1}{c||}{L} \\
\hhline{|:=::=::=====::=::=====:|}
$\phantom{-20 < \mathord{}} p > 20\phantom{-}$ & 52.7 & 0.3 & --- & --- & 52.7 & --- & 48.5 & 0.3 & --- & --- & 48.5 & --- \\
\hhline{||-||-||-|-|-|-|-||-||-|-|-|-|-||}
$\phantom{-}10 < p \leq 20\phantom{-}$ & 11.7 & --- & --- & --- & 11.7 & --- & 16.0 & --- & --- & --- & 16.0 & --- \\
\hhline{||-||-||-|-|-|-|-||-||-|-|-|-|-||}
$\phantom{-1}5 < p \leq 10\phantom{-}$ & 5.4 & --- & --- & --- & 5.4 & --- & 7.5 & --- & --- & --- & 7.5 & --- \\
\hhline{||-||-||-|-|-|-|-||-||-|-|-|-|-||}
$\phantom{-1}2 < p \leq 5\phantom{-0}$ & 2.4 & --- & --- & --- & 2.4 & --- & 1.8 & --- & --- & --- & 1.8 & --- \\
\hhline{||-||-||-|-|-|-|-||-||-|-|-|-|-||}
$\phantom{-1}0 < p \leq 2\phantom{-0}$ & 0.3 & --- & --- & --- & 0.3 & 1.5 & 0.6 & --- & --- & --- & 0.6 & 1.5 \\
\hhline{|:=::=::=:=:=:=:=::=::=:=:=:=:=:|}
same precision & 24.1 & 96.4 & 96.7 & 96.7 & 24.1 & 95.2 & 19.0 & 93.1 & 93.4 & 93.4 & 19.0 & 91.9 \\
\hhline{||-||-||-|-|-|-|-||-||-|-|-|-|-||}
unknown & 3.3 & 3.3 & 3.3 & 3.3 & 3.3 & 3.3 & 6.6 & 6.6 & 6.6 & 6.6 & 6.6 & 6.6 \\
\hhline{|b:=:b:=:b:=====:b:=:b:=====:b|}

\multicolumn{13}{l}{}\\

\hhline{~|t:======:t:======:t|}
    \multicolumn{1}{l||}{Goal Dep.}
  & \multicolumn{6}{c||}{without Struct Info}
  & \multicolumn{6}{c||}{with Struct Info} \\
\hhline{|t:=::=:t:=====::=:t:=====:|}
 Prec. class
 & \multicolumn{1}{c||}{O}
 & \multicolumn{1}{c|}{I}
 & \multicolumn{1}{c|}{G}
 & \multicolumn{1}{c|}{F}
 & \multicolumn{1}{c|}{GF}
 & \multicolumn{1}{c||}{L}
 & \multicolumn{1}{c||}{O}
 & \multicolumn{1}{c|}{I}
 & \multicolumn{1}{c|}{G}
 & \multicolumn{1}{c|}{F}
 & \multicolumn{1}{c|}{GF}
 & \multicolumn{1}{c||}{L} \\
\hhline{|:=::=::=====::=::=====:|}
$\phantom{-20 < \mathord{}} p > 20\phantom{-}$ & 5.9 & --- & --- & --- & 5.9 & --- & 5.9 & --- & --- & --- & 5.9 & --- \\
\hhline{||-||-||-|-|-|-|-||-||-|-|-|-|-||}
$\phantom{-}10 < p \leq 20\phantom{-}$ & 4.5 & --- & --- & --- & 4.5 & --- & 5.4 & --- & --- & --- & 5.4 & --- \\
\hhline{||-||-||-|-|-|-|-||-||-|-|-|-|-||}
$\phantom{-1}5 < p \leq 10\phantom{-}$ & 7.7 & 0.5 & --- & --- & 7.7 & --- & 5.4 & 0.5 & --- & --- & 5.4 & --- \\
\hhline{||-||-||-|-|-|-|-||-||-|-|-|-|-||}
$\phantom{-1}2 < p \leq 5\phantom{-0}$ & 13.1 & --- & --- & --- & 13.1 & --- & 12.2 & --- & --- & --- & 12.2 & --- \\
\hhline{||-||-||-|-|-|-|-||-||-|-|-|-|-||}
$\phantom{-1}0 < p \leq 2\phantom{-0}$ & 8.1 & --- & --- & --- & 8.1 & 0.5 & 10.0 & --- & --- & --- & 10.0 & --- \\
\hhline{|:=::=::=:=:=:=:=::=::=:=:=:=:=:|}
same precision & 57.0 & 95.9 & 96.4 & 96.4 & 57.0 & 95.9 & 51.6 & 90.0 & 90.5 & 90.5 & 51.6 & 90.5 \\
\hhline{||-||-||-|-|-|-|-||-||-|-|-|-|-||}
unknown & 3.6 & 3.6 & 3.6 & 3.6 & 3.6 & 3.6 & 9.5 & 9.5 & 9.5 & 9.5 & 9.5 & 9.5 \\
\hhline{|b:=:b:=:b:=====:b:=:b:=====:b|}
\end{tabular}

\bigskip\bigskip

\begin{tabular}{||l||r|r||r|r||}
\hhline{|t:=:t:==:t:==:t|}
Time diff. class &
\multicolumn{2}{c||}{Goal Ind.} &
\multicolumn{2}{c||}{Goal Dep.} \\
\hhline{|b:=::==::==:|}
\multicolumn{1}{c||}{} &
\multicolumn{1}{c|}{w/o SI} &
\multicolumn{1}{c||}{with SI} &
\multicolumn{1}{c|}{w/o SI} &
\multicolumn{1}{c||}{with SI} \\
\hhline{|t:=::==::==:|}
$\phantom{0.5 < \mathord{}} \text{degradation} > 1$ & --- & 0.6 & --- & 0.9 \\
 \hhline{||-||-|-||-|-||}
$0.5 < \text{degradation} \leq 1$ & 0.3 & --- & 0.5 & --- \\
 \hhline{||-||-|-||-|-||}
$0.2 < \text{degradation} \leq 0.5$ & --- & 0.6 & 0.5 & 1.4 \\
 \hhline{||-||-|-||-|-||}
$0.1 < \text{degradation} \leq 0.2$ & 0.3 & --- & --- & 0.5 \\
 \hhline{|:=::=:=::=:=:|}
$\phantom{0.5 < \mathord{}} \text{both timed out}$ & 3.3 & 6.6 & 3.6 & 9.5 \\
 \hhline{||-||-|-||-|-||}
$\phantom{0.5 < \mathord{}} \text{same time}$ & 88.6 & 85.2 & 87.3 & 82.8 \\
 \hhline{|:=::=:=::=:=:|}
$0.1 < \text{improvement} \leq 0.2$ & 1.2 & 1.2 & 1.8 & 1.4 \\
 \hhline{||-||-|-||-|-||}
$0.2 < \text{improvement} \leq 0.5$ & 2.4 & 2.4 & 1.8 & 0.9 \\
 \hhline{||-||-|-||-|-||}
$0.5 < \text{improvement} \leq 1$ & 2.1 & 0.9 & 2.3 & 0.9 \\
 \hhline{||-||-|-||-|-||}
$\phantom{0.5 < \mathord{}} \text{improvement} > 1$ & 1.8 & 2.4 & 2.3 & 1.8 \\
\hhline{|b:=:b:==:b:==:b|}
\end{tabular}

\bigskip\bigskip

\begin{tabular}{||c||r|r|r||r|r|r||r|r|r||r|r|r||}
\hhline{|t:=:t:======:t:======:t|}
\multicolumn{1}{||c||}{Total time class}
 & \multicolumn{6}{c||}{Goal Independent}
 & \multicolumn{6}{c||}{Goal Dependent} \\
\hhline{|b:=::===:t:===::===:t:===:|}
\multicolumn{1}{c||}{}
 & \multicolumn{3}{c||}{without SI}
 & \multicolumn{3}{c||}{with SI}
 & \multicolumn{3}{c||}{without SI}
 & \multicolumn{3}{c||}{with SI} \\
\hhline{~|:===::===::===::===:|}
\multicolumn{1}{c||}{}
 & \multicolumn{1}{c|}{\%1} & \multicolumn{1}{c|}{\%2}
 & \multicolumn{1}{c||}{$\Delta$} & \multicolumn{1}{c|}{\%1}
 & \multicolumn{1}{c|}{\%2} & \multicolumn{1}{c||}{$\Delta$}
 & \multicolumn{1}{c|}{\%1} & \multicolumn{1}{c|}{\%2}
 & \multicolumn{1}{c||}{$\Delta$} & \multicolumn{1}{c|}{\%1}
 & \multicolumn{1}{c|}{\%2} & \multicolumn{1}{c||}{$\Delta$} \\
\hhline{|t:=::===::===::===::===:|}
$\text{timed out}$ & 3.3 & 3.3 & --- & 6.6 & 6.6 & --- & 3.6 & 3.6 & --- & 9.5 & 9.5 & --- \\
 \hhline{||-||-|-|-||-|-|-||-|-|-||-|-|-||}
$\phantom{0.5 < \mathord{}} t > 10\phantom{.}$ & 9.0 & 9.0 & --- & 8.4 & 8.4 & --- & 7.2 & 7.2 & --- & 8.6 & 8.6 & --- \\
 \hhline{||-||-|-|-||-|-|-||-|-|-||-|-|-||}
$\phantom{0.}5 < t \leq 10\phantom{.}$ & 0.3 & 0.3 & --- & 1.5 & 1.5 & --- & 1.4 & 1.4 & --- & 1.8 & 1.8 & --- \\
 \hhline{||-||-|-|-||-|-|-||-|-|-||-|-|-||}
$\phantom{0.}1 < t \leq 5\phantom{.5}$ & 7.5 & 7.5 & --- & 6.6 & 6.6 & --- & 3.6 & 3.6 & --- & 5.0 & 5.0 & --- \\
 \hhline{||-||-|-|-||-|-|-||-|-|-||-|-|-||}
$0.5 < t \leq 1\phantom{.5}$ & 2.7 & 2.7 & --- & 3.3 & 3.6 & 0.3 & 5.4 & 5.9 & 0.5 & 3.2 & 3.2 & --- \\
 \hhline{||-||-|-|-||-|-|-||-|-|-||-|-|-||}
$0.2 < t \leq 0.5$ & 8.4 & 8.7 & 0.3 & 10.2 & 10.5 & 0.3 & 13.1 & 12.7 & -0.5 & 13.6 & 14.0 & 0.5 \\
 \hhline{||-||-|-|-||-|-|-||-|-|-||-|-|-||}
$\phantom{0.5 < \mathord{}} t \leq 0.2$ & 68.7 & 68.4 & -0.3 & 63.3 & 62.7 & -0.6 & 65.6 & 65.6 & --- & 58.4 & 57.9 & -0.5 \\
\hhline{|b:=:b:===:b:===:b:===:b:===:b|}
\end{tabular}

\caption{$Pos \times \PSDFL$ versus $\Pos \times \PSDGFL$.}
\label{tab:ground-or-free}
\end{table*}

To summarize, the incorporation of the set of ground-or-free
variables is cheap, both in terms of computational complexity and
in terms of code to be written.
As far as computational complexity is concerned this extension
looks particularly promising, since the possibility of avoiding
star-unions has the potential of absorbing its overhead
if not of giving rise to a speed-up.

Thus the domain $\Pos \times \SGFL$ was experimentally evaluated
on our benchmark suite, with and without the structural
information provided by $\Pattern(\cdot)$,
both in a goal-dependent and in a goal-independent way,
and the results compared with those previously obtained
for the domain $\Pos \times \SFL$.
Note that the implementation uses the non-redundant version
$\PSDGFL \defeq \PSD \times F \times \GF \times L$.
In the precision comparisons of Table~\ref{tab:ground-or-free},
the new column labeled GF reports precision improvements
measured on the ground-or-free property itself.\footnote{For this comparison,
in the analysis using $\Pos \times \SFL$, the number of ground-or-free
variables is computed by summing the number of ground variables
with the number of free variables.}

As far as the timings are concerned, the experimentation
fully confirms our qualitative reasoning:
efficiency improvements are more frequent than degradations and,
even with widening operators switched off, the distributions of
the total analysis times show minor changes only.
As for precision,
disregarding the many improvements in the GF columns,
few changes can be observed, and almost all of these concern just the
linearity information.\footnote{%
In fact the sole improvement to the number of independent pairs
is due to a synthetic benchmark, named \texttt{gof},
that was explicitly written to show
that variable independence could be affected.}

The results in Table~\ref{tab:ground-or-free},
show that tracking ground-or-free variables,
while being potentially useful for improving the precision
of a sharing analysis, rarely reaches such a goal.
In contrast, the precision gains on the ground-or-free property
itself are remarkable,
affecting from 39\% to 74\% of the programs in the benchmark
suite.
It is possible to foresee several \emph{direct} applications
for this information that, together with the just mentioned negligible
computational cost, fully justify the inclusion
of this enhancement in a static analyzer.
In particular, there are at least two ways in which a knowledge of
ground-or-free variables could improve the concrete unification procedure.

The first case applies in the context of
occurs-check reduction~\cite{Sondergaard86,CrnogoracKS96},
that is when a program designed for a logic programming system
performing the occurs-check is to be run on top of a system
omitting this test.
In order to ensure correct execution,
all the explicit and implicit unifications in the program
are treated as if the ISO Prolog built-in
\texttt{unify\_with\_occurs\_check/2}
was used to perform them.
In order to minimize the performance overhead, it is important
to detect, as precisely as possible and at compile-time,
those \emph{NSTO}
(short for \emph{Not Subject To the Occurs-check}
\cite{DeransartFT91,ISO-Prolog-part-1}) unifications
where the occurs-check will not be needed.
For these unifications, \texttt{=/2} can safely be used;
for the remaining ones, the program will have to be transformed
so that \texttt{unify\_with\_occurs\_check/2} is explicitly called
to perform them.
Ground-or-freeness can be of help for this application,
since a unification between two ground-or-free variables is \emph{NSTO}.
Note that this is an improvement
with respect to the technique used in \cite{CrnogoracKS96}, since
it is not required that the two considered variables are independent.

As a second application, ground-or-freeness can be useful to replace
the full concrete unification procedure by a simplified version.
Since a ground-or-free term is either ground or free,
a \emph{single} run-time test for freeness will discriminate between
the two cases:
if this test succeeds, unification can be implemented by
a single assignment;
if the test fails, any specialized code for unification
with a ground term can be safely invoked.
In particular, when unifying two ground-or-free variables that are
not free at run-time, the full unification procedure can be replaced
by a simpler recursive test for equivalence.

\section{More Precise Exploitation of Linearity}
\label{sec:enhancedlin}

In \cite{King94}, A.~King proposes a domain for sharing analysis
that performs a quite precise tracking of linearity.
Roughly speaking, each sharing group in a sharing-set carries
its own linearity information.
In contrast, in the approach of \cite{Langen90th}, which is the
one usually followed, a set of definitely linear variables
is recorded along with each sharing-set.
The proposal in \cite{King94} gives rise to a domain that is
quite different from the ones presented here.
Since \cite{King94} does not provide an experimental evaluation
and we are unaware of any subsequent work on the subject,
the question whether this more precise tracking of linearity
is actually worthwhile (both in terms of precision and efficiency)
seems open.

\begin{table*}
\centering
\begin{tabular}{||c||r||r|r|r|r||r||r|r|r|r||}
\hhline{~|t:=====:t:=====:t|}
    \multicolumn{1}{c||}{}
  & \multicolumn{5}{c||}{Goal Independent}
  & \multicolumn{5}{c||}{Goal Dependent} \\
\hhline{|t:=::=:t:====::=:t:====:|}
 Prec. class
 & \multicolumn{1}{c||}{O}
 & \multicolumn{1}{c|}{I}
 & \multicolumn{1}{c|}{G}
 & \multicolumn{1}{c|}{F}
 & \multicolumn{1}{c||}{L}
 & \multicolumn{1}{c||}{O}
 & \multicolumn{1}{c|}{I}
 & \multicolumn{1}{c|}{G}
 & \multicolumn{1}{c|}{F}
 & \multicolumn{1}{c||}{L} \\
\hhline{|:=::=::====::=::====:|}
$\phantom{-20 < \mathord{}} p > 20\phantom{-}$ & 0.3 & 0.3 & --- & --- & --- & --- & --- & --- & --- & --- \\
\hhline{||-||-||-|-|-|-||-||-|-|-|-||}
$\phantom{-1}2 < p \leq 5\phantom{-0}$ & --- & --- & --- & --- & --- & 0.5 & 0.5 & --- & --- & --- \\
\hhline{|:=::=::=:=:=:=::=::=:=:=:=:|}
same precision & 93.1 & 93.1 & 93.4 & 93.4 & 93.4 & 90.0 & 90.0 & 90.5 & 90.5 & 90.5 \\
\hhline{||-||-||-|-|-|-||-||-|-|-|-||}
unknown & 6.6 & 6.6 & 6.6 & 6.6 & 6.6 & 9.5 & 9.5 & 9.5 & 9.5 & 9.5 \\
\hhline{|b:=:b:=:b:====:b:=:b:====:b|}
\end{tabular}

\bigskip\bigskip

\begin{tabular}{||l||r|r||}
\hhline{|t:=:t:==:t|}
\multicolumn{1}{||c||}{Time difference class} &
\multicolumn{2}{c||}{\% benchmarks} \\
\hhline{|b:=::==:|}
\multicolumn{1}{c||}{} &
\multicolumn{1}{c|}{Goal Ind.} &
\multicolumn{1}{c||}{Goal Dep.} \\
\hhline{|t:=::==:|}
$\phantom{0.5 < \mathord{}} \text{degradation} > 1$ & 0.3 & --- \\
 \hhline{||-||-|-||}
$0.5 < \text{degradation} \leq 1$ & --- & --- \\
 \hhline{||-||-|-||}
$0.2 < \text{degradation} \leq 0.5$ & --- & --- \\
 \hhline{||-||-|-||}
$0.1 < \text{degradation} \leq 0.2$ & 0.3 & 0.5 \\
 \hhline{|:=::=:=:|}
$\phantom{0.5 < \mathord{}} \text{both timed out}$ & 6.6 & 9.5 \\
 \hhline{||-||-|-||}
$\phantom{0.5 < \mathord{}} \text{same time}$ & 85.2 & 83.7 \\
 \hhline{|:=::=:=:|}
$0.1 < \text{improvement} \leq 0.2$ & 0.9 & 1.8 \\
 \hhline{||-||-|-||}
$0.2 < \text{improvement} \leq 0.5$ & 2.4 & 0.5 \\
 \hhline{||-||-|-||}
$0.5 < \text{improvement} \leq 1$ & 0.6 & 2.7 \\
 \hhline{||-||-|-||}
$\phantom{0.5 < \mathord{}} \text{improvement} > 1$ & 3.6 & 1.4 \\
\hhline{|b:=:b:==:b|}
\end{tabular}

\bigskip\bigskip

\begin{tabular}{||c||r|r|r||r|r|r||}
\hhline{|t:=:t:===:t:===:t|}
\multicolumn{1}{||c||}{Total time class}
 & \multicolumn{3}{c||}{Goal Ind.}
 & \multicolumn{3}{c||}{Goal Dep.} \\
\hhline{|b:=::===::===:|}
\multicolumn{1}{c||}{}
 & \multicolumn{1}{c|}{\%1} & \multicolumn{1}{c|}{\%2}
 & \multicolumn{1}{c||}{$\Delta$} & \multicolumn{1}{c|}{\%1}
 & \multicolumn{1}{c|}{\%2} & \multicolumn{1}{c||}{$\Delta$} \\
\hhline{|t:=::===::===:|}
$\text{timed out}$ & 6.6 & 6.6 & --- & 9.5 & 9.5 & --- \\
 \hhline{||-||-|-|-||-|-|-||}
$\phantom{0.5 < \mathord{}} t > 10\phantom{.}$ & 8.4 & 8.4 & --- & 8.6 & 8.6 & --- \\
 \hhline{||-||-|-|-||-|-|-||}
$\phantom{0.}5 < t \leq 10\phantom{.}$ & 1.5 & 1.5 & --- & 1.8 & 1.8 & --- \\
 \hhline{||-||-|-|-||-|-|-||}
$\phantom{0.}1 < t \leq 5\phantom{.5}$ & 6.6 & 6.6 & --- & 5.0 & 5.0 & --- \\
 \hhline{||-||-|-|-||-|-|-||}
$0.5 < t \leq 1\phantom{.5}$ & 3.3 & 3.3 & --- & 3.2 & 3.2 & --- \\
 \hhline{||-||-|-|-||-|-|-||}
$0.2 < t \leq 0.5$ & 10.2 & 11.1 & 0.9 & 13.6 & 14.0 & 0.5 \\
 \hhline{||-||-|-|-||-|-|-||}
$\phantom{0.5 < \mathord{}} t \leq 0.2$ & 63.3 & 62.3 & -0.9 & 58.4 & 57.9 & -0.5 \\
\hhline{|b:=:b:===:b:===:b|}
\end{tabular}

\caption{The effect of enhanced linearity on $\Pattern(\Pos \times \PSDFL)$.}
\label{tab:simplepos-struct-vs-elin-struct}
\end{table*}

What interests us here is that part of the theoretical work presented
in \cite{King94} may be usefully applied even in the more classical
treatments of linearity such as the one being used in this paper.
As far as we can tell, this fact was first noted in~\cite{BagnaraZH00}.

In \cite{King94}, point 3 of Lemma 5
(which is reported to be proven in \cite{King93TR})
states that, if $s$ is a linear term independent from a term $t$,
then in the unifier for $s = t$ any sharing
between the variables in $s$ is necessarily caused by those
variables that can occur more than once in $t$.

This result can be exploited even when using
the domain $\SFL$.
Given the abstract element $\sfl = \langle \sh, f, l \rangle$,
let $x \in (l \setdiff f)$ be a non-free but linear variable
and let $t$ be a non-linear term such that
$\ind_{\sfl}(x,t)$.
Let also $V_x$, $V_t$, $V_{xt}$, $R_x$ and $R_t$ be as given
in Definition~\ref{def:abs-funcs-SFL}.
In such a situation, when abstractly evaluating the binding $x = t$,
the standard $\amgu$ operator gives the set-sharing component
\[
  \sh' = \irel(V_{xt}, \sh) \union \bin(R_x^\star, R_t).
\]
Suppose the set $V_t$ is partitioned into the two components
$V_{t}^\mathrm{l}$ and $V_{t}^\mathrm{nl}$,
where $V_{t}^\mathrm{nl}$ is the set of the ``problematic'' variables,
that is, those variables that potentially make $t$ a non-linear term.
Formally,
\begin{align*}
  V_{t}^\mathrm{l}
    &\defeq
      \sset{
        y \in \vars(t)
      }{
        y \in l \\
        y \Min \mvars(t) \implies y \notin \vars(\sh) \\
        \forall z \in \vars(t) \itc \bigl( y = z \lor \ind_{\sfl}(y, z)\bigr)
      }; \\
  V_{t}^\mathrm{nl} &\defeq V_t \setdiff V_{t}^\mathrm{l}.
\end{align*}
Let $R_{t}^\mathrm{l} = \rel(V_{t}^\mathrm{l}, \sh)$
and $R_{t}^\mathrm{nl} = \rel(V_{t}^\mathrm{nl}, \sh)$.
Note that $R_{t}^\mathrm{nl} \neq \emptyset$,
because $t$ is a non-linear term.
If also $R_{t}^\mathrm{l} \neq \emptyset$
then the standard $\amgu$ can be replaced by an improved version
(denoted by $\amgu_k$) computing the following set-sharing component:
\begin{align*}
  \sh'_k &= \irel(V_{xt}, \sh)
              \union
            \bin(R_x, R_{t}^\mathrm{l})
              \union
            \bin(R_x^\star, R_{t}^\mathrm{nl}).
\end{align*}
As a consequence of King's result \cite[Lemma 5]{King94},
only $R_{t}^\mathrm{nl}$
(the relevant component of $\sh$ with respect to
the problematic variables $V_{t}^\mathrm{nl}$)
has to be combined with $R_x^\star$
while $R_{t}^\mathrm{l}$ can be combined with just
$R_x$ (without the star-union).

For a working example,
suppose $\VI = \{v,w,x,y,z\}$ is the set of variables of interest
and consider the $\SFL$ element
\begin{align*}
  \sfl &\defeq
         \bigl\langle
           \{ vx, wx, y, z \},
           \{ v, w, y \},
           \{ v, w, x, y \}
         \bigr\rangle
\intertext{%
with the binding $x = f(y,z)$.
Note that all the applicability conditions specified above are met:
in particular $t = f(y, z)$ is not linear because $z \notin l$.
As $R_x = \{ vx, wx \}$ and $R_t = \{ y, z \}$,
a standard analysis would compute
}
  \sfl' &= \amgu\bigl(\sfl, x = f(y,z)\bigr) \\
        &= \bigl\langle
             \{ vwxy, vwxz, vxy, vxz, wxy, wxz \},
             \emptyset,
             \{ y \}
           \bigr\rangle. \\
\intertext{%
On the other hand, since
$V_t^\mathrm{l} = \{ y \}$ and $V_t^\mathrm{nl} = \{ z \}$,
the enhanced analysis would compute
}
  \sfl'_k & = \amgu_k\bigl(\sfl, x = f(y,z)\bigr) \\
          &= \bigl\langle
               \{ vwxz, vxy, vxz, wxy, wxz \},
               \emptyset,
               \{ y \}
             \bigr\rangle.
\end{align*}
Note that
$\sfl'_k$ does not include the sharing group $vwxy$.
This means that, if in the sequel of the computation
variable $z$ is bound to a ground term,
then variables $v$ and $w$ will be known to be definitely independent.
This independence
 is not captured when using the standard $\amgu$
since $\sfl'$ includes the sharing group $vwxy$,
and therefore the variables $v$ and $w$
will potentially share even after grounding $z$.

The experimental evaluation for this enhancement is reported
in Table~\ref{tab:simplepos-struct-vs-elin-struct}.
The comparison of times shows that the efficiency of the analysis,
when affected, is more likely to be improved than degraded.
As for the precision, improvements are observed for only two programs;
moreover, these are synthetic benchmarks such as the above example.
Nevertheless, despite its limited practical relevance,
this result
demonstrates that the standard combination
of $\Sharing$ with linearity information is \emph{not} optimal,
even when all the possible orderings
of the non-grounding bindings are tried.

\section{Sharing and Freeness}
\label{sec:enhancedfree}

As noted by several authors \cite{BruynoogheCM94,BuenodlBH94,CabezaH94},
the standard combination of $\Sharing$ and $\Free$ is not optimal.
G.~Fil\'e \cite{File94} formally identified the reduced product
of these domains and proposed an improved abstract unification operator.
This new operator exploits two properties that hold
for the most precise abstract description of a \emph{single}
concrete substitution:
\begin{enumerate}
\item
each free variable occurs in exactly one sharing group;
\item
two free variables occur in the same sharing group if and only if they
are aliases (i.e., they have become the same variable).
\end{enumerate}
When considering the general case,
where sets of concrete substitutions come into play,
property 1 can be used to (partially) recover disjunctive information.
In particular, it is possible to decompose an abstract description
into a set of (maximal) descriptions that necessarily come from
different computation paths, each one satisfying property 1.
The abstract unification procedure can thus be computed separately on each
component, and the results of each subcomputation are then joined
to give the final description.
As such components are more precise than the original description
(they possibly contain more ground variables and less sharing pairs),
precision gains can be obtained.

Furthermore, by exploiting property 2 on each component,
it is possible to correctly infer that for some of them
the computation will fail due to a functor clash
(or to the occurs-check, if considering a system working on finite trees).
Note that a similar improvement is possible even without decomposing
the abstract description.
As an example, consider an abstract element such as the following:
\[
  \sfl = \bigl\langle \{xy, u, v\}, \{x, y\}, \{x, y\} \bigr\rangle.
\]
Since the sharing group $xy$ is the only one where the free variables
$x$ and $y$ occur, property 2 states that $x$ and $y$ are indeed
the same variable in all the concrete computation states
described by $\sfl \in \SFL$.
Therefore, when abstractly evaluating the substitution
$\bigl\{ x = f(u), y =  g(v) \bigr\}$,
it can be safely concluded that its concrete counterparts
will result in failure due to the functor clash.
In the same circumstances, it can also be concluded that
a concrete substitution corresponding to, say,
$\bigl\{ x = f(y) \bigr\}$
will cause a failure of the occurs-check, if this is performed.

\begin{table*}
\centering
\begin{tabular}{||c||r||r|r|r|r||r||r|r|r|r||}
\hhline{~|t:=====:t:=====:t|}
    \multicolumn{1}{l||}{Goal Independent}
  & \multicolumn{5}{c||}{without Struct Info}
  & \multicolumn{5}{c||}{with Struct Info} \\
\hhline{|t:=::=:t:====::=:t:====:|}
 Prec. class
 & \multicolumn{1}{c||}{O}
 & \multicolumn{1}{c|}{I}
 & \multicolumn{1}{c|}{G}
 & \multicolumn{1}{c|}{F}
 & \multicolumn{1}{c||}{L}
 & \multicolumn{1}{c||}{O}
 & \multicolumn{1}{c|}{I}
 & \multicolumn{1}{c|}{G}
 & \multicolumn{1}{c|}{F}
 & \multicolumn{1}{c||}{L} \\
\hhline{|:=::=::====::=::====:|}
$\phantom{-20 < \mathord{}} p > 20\phantom{-}$ & 0.3 & 0.3 & --- & --- & --- & --- & --- & --- & --- & --- \\
\hhline{||-||-||-|-|-|-||-||-|-|-|-||}
$\phantom{-1}5 < p \leq 10\phantom{-}$ & --- & --- & --- & --- & --- & 0.3 & --- & --- & --- & 0.3 \\
\hhline{||-||-||-|-|-|-||-||-|-|-|-||}
$\phantom{-1}0 < p \leq 2\phantom{-0}$ & 0.9 & 0.3 & --- & --- & 0.6 & 3.6 & 3.0 & --- & --- & 0.6 \\
\hhline{|:=::=::=:=:=:=::=::=:=:=:=:|}
same precision & 94.6 & 95.2 & 95.8 & 95.8 & 95.2 & 86.1 & 87.0 & 90.1 & 90.1 & 89.2 \\
\hhline{||-||-||-|-|-|-||-||-|-|-|-||}
unknown & 4.2 & 4.2 & 4.2 & 4.2 & 4.2 & 9.9 & 9.9 & 9.9 & 9.9 & 9.9 \\
\hhline{|b:=:b:=:b:====:b:=:b:====:b|}

\multicolumn{11}{l}{}\\
\multicolumn{11}{l}{}\\

\hhline{~|t:=====:t:=====:t|}
    \multicolumn{1}{l||}{Goal Dependent}
  & \multicolumn{5}{c||}{without Struct Info}
  & \multicolumn{5}{c||}{with Struct Info} \\
\hhline{|t:=::=:t:====::=:t:====:|}
 Prec. class
 & \multicolumn{1}{c||}{O}
 & \multicolumn{1}{c|}{I}
 & \multicolumn{1}{c|}{G}
 & \multicolumn{1}{c|}{F}
 & \multicolumn{1}{c||}{L}
 & \multicolumn{1}{c||}{O}
 & \multicolumn{1}{c|}{I}
 & \multicolumn{1}{c|}{G}
 & \multicolumn{1}{c|}{F}
 & \multicolumn{1}{c||}{L} \\
\hhline{|:=::=::====::=::====:|}
same precision & 96.4 & 96.4 & 96.4 & 96.4 & 96.4 & 89.6 & 89.6 & 89.6 & 89.6 & 89.6 \\
\hhline{||-||-||-|-|-|-||-||-|-|-|-||}
unknown & 3.6 & 3.6 & 3.6 & 3.6 & 3.6 & 10.4 & 10.4 & 10.4 & 10.4 & 10.4 \\
\hhline{|b:=:b:=:b:====:b:=:b:====:b|}
\end{tabular}

\bigskip\bigskip

\begin{tabular}{||l||r|r||r|r||}
\hhline{|t:=:t:==:t:==:t|}
Time diff. class &
\multicolumn{2}{c||}{Goal Ind.} &
\multicolumn{2}{c||}{Goal Dep.} \\
\hhline{|b:=::==::==:|}
\multicolumn{1}{c||}{} &
\multicolumn{1}{c|}{w/o SI} &
\multicolumn{1}{c||}{with SI} &
\multicolumn{1}{c|}{w/o SI} &
\multicolumn{1}{c||}{with SI} \\
\hhline{|t:=::==::==:|}
$\phantom{0.5 < \mathord{}} \text{degradation} > 1$ & 9.6 & 13.6 & 3.2 & 5.9 \\
 \hhline{||-||-|-||-|-||}
$0.5 < \text{degradation} \leq 1$ & 0.6 & 1.8 & 1.4 & 1.4 \\
 \hhline{||-||-|-||-|-||}
$0.2 < \text{degradation} \leq 0.5$ & 3.3 & 2.4 & 1.8 & 3.6 \\
 \hhline{||-||-|-||-|-||}
$0.1 < \text{degradation} \leq 0.2$ & 0.6 & 1.5 & 2.3 & 1.4 \\
 \hhline{|:=::=:=::=:=:|}
$\phantom{0.5 < \mathord{}} \text{both timed out}$ & 3.3 & 6.6 & 3.6 & 9.5 \\
 \hhline{||-||-|-||-|-||}
$\phantom{0.5 < \mathord{}} \text{same time}$ & 82.2 & 73.5 & 87.8 & 77.8 \\
 \hhline{|:=::=:=::=:=:|}
$0.1 < \text{improvement} \leq 0.2$ & --- & --- & --- & --- \\
 \hhline{||-||-|-||-|-||}
$0.2 < \text{improvement} \leq 0.5$ & 0.3 & --- & --- & --- \\
 \hhline{||-||-|-||-|-||}
$0.5 < \text{improvement} \leq 1$ & --- & --- & --- & --- \\
 \hhline{||-||-|-||-|-||}
$\phantom{0.5 < \mathord{}} \text{improvement} > 1$ & --- & 0.6 & --- & 0.5 \\
\hhline{|b:=:b:==:b:==:b|}
\end{tabular}

\bigskip\bigskip

\begin{tabular}{||c||r|r|r||r|r|r||r|r|r||r|r|r||}
\hhline{|t:=:t:======:t:======:t|}
\multicolumn{1}{||c||}{Total time class}
 & \multicolumn{6}{c||}{Goal Independent}
 & \multicolumn{6}{c||}{Goal Dependent} \\
\hhline{|b:=::===:t:===::===:t:===:|}
\multicolumn{1}{c||}{}
 & \multicolumn{3}{c||}{without SI}
 & \multicolumn{3}{c||}{with SI}
 & \multicolumn{3}{c||}{without SI}
 & \multicolumn{3}{c||}{with SI} \\
\hhline{~|:===::===::===::===:|}
\multicolumn{1}{c||}{}
 & \multicolumn{1}{c|}{\%1} & \multicolumn{1}{c|}{\%2}
 & \multicolumn{1}{c||}{$\Delta$} & \multicolumn{1}{c|}{\%1}
 & \multicolumn{1}{c|}{\%2} & \multicolumn{1}{c||}{$\Delta$}
 & \multicolumn{1}{c|}{\%1} & \multicolumn{1}{c|}{\%2}
 & \multicolumn{1}{c||}{$\Delta$} & \multicolumn{1}{c|}{\%1}
 & \multicolumn{1}{c|}{\%2} & \multicolumn{1}{c||}{$\Delta$} \\
\hhline{|t:=::===::===::===::===:|}
$\text{timed out}$ & 3.3 & 4.2 & 0.9 & 6.6 & 9.9 & 3.3 & 3.6 & 3.6 & --- & 9.5 & 10.4 & 0.9 \\
 \hhline{||-||-|-|-||-|-|-||-|-|-||-|-|-||}
$\phantom{0.5 < \mathord{}} t > 10\phantom{.}$ & 9.0 & 9.6 & 0.6 & 8.4 & 8.4 & --- & 7.2 & 7.2 & --- & 8.6 & 8.1 & -0.5 \\
 \hhline{||-||-|-|-||-|-|-||-|-|-||-|-|-||}
$\phantom{0.}5 < t \leq 10\phantom{.}$ & 0.3 & 0.9 & 0.6 & 1.5 & 1.2 & -0.3 & 1.4 & 1.4 & --- & 1.8 & 1.8 & --- \\
 \hhline{||-||-|-|-||-|-|-||-|-|-||-|-|-||}
$\phantom{0.}1 < t \leq 5\phantom{.5}$ & 7.5 & 6.9 & -0.6 & 6.6 & 5.7 & -0.9 & 3.6 & 3.6 & --- & 5.0 & 4.5 & -0.5 \\
 \hhline{||-||-|-|-||-|-|-||-|-|-||-|-|-||}
$0.5 < t \leq 1\phantom{.5}$ & 2.7 & 2.1 & -0.6 & 3.3 & 4.5 & 1.2 & 5.4 & 5.9 & 0.5 & 3.2 & 3.2 & --- \\
 \hhline{||-||-|-|-||-|-|-||-|-|-||-|-|-||}
$0.2 < t \leq 0.5$ & 8.4 & 8.4 & --- & 10.2 & 12.0 & 1.8 & 13.1 & 12.7 & -0.5 & 13.6 & 14.9 & 1.4 \\
 \hhline{||-||-|-|-||-|-|-||-|-|-||-|-|-||}
$\phantom{0.5 < \mathord{}} t \leq 0.2$ & 68.7 & 67.8 & -0.9 & 63.3 & 58.1 & -5.1 & 65.6 & 65.6 & --- & 58.4 & 57.0 & -1.4 \\
\hhline{|b:=:b:===:b:===:b:===:b:===:b|}
\end{tabular}

\caption{The effect of enhanced freeness on $\Pos \times \PSDFL$.}
\label{tab:simplepos-vs-efree}
\end{table*}

As was the case for the reduced product between $\Pos$ and $\SH$
(see Section~\ref{sec:enhancedpos}),
the interaction between the enhanced abstract unification operator
and the elimination of $\rho$-redundant elements
can lead to results that are not correct.

To see this, let $\VI = \{w, x, y, z\}$ and consider the set of concrete
substitutions $\Sigma = \wp(\sigma)$,
where $\sigma = \{ x \mapsto v, y \mapsto v, z \mapsto v \}$
(note that $v \notin \VI$).
The abstract element describing $\Sigma$ is
\(
  \sfl = \langle \sh, f, l \rangle \in \SFL
\),
where
$\sh = \{ w, x, xy, xyz, xz, y, yz, z \}$ and
$f = l = \VI$.
Suppose that the implementation represents $\sfl$
by using the reduced element
$\sfl_\reduced = \langle \sh_\reduced, f, l \rangle$,
where $\sh_\reduced = \sh \setdiff \{ xyz \}$,
so that $\sh = \rho(\sh_\reduced)$.

According to the specification of the enhanced operator,
$\sfl_\reduced$ can be decomposed into
the following four components:
\begin{align*}
  c_1 &= \bigl\langle \{ w, x, y, z \}, f, l \bigr\rangle, 
&
  c_3 &= \bigl\langle \{ w, xz, y \}, f, l \bigr\rangle, \\
  c_2 &= \bigl\langle \{ w, x, yz \}, f, l \bigr\rangle, 
&
  c_4 &= \bigl\langle \{ w, xy, z \}, f, l \bigr\rangle.
\end{align*}

Consider the binding $x = f(y, w)$
and, for each $i \in \{1, \ldots, 4\}$,
the computation of
\(
  c'_i = \langle \sh'_i, f'_i, l'_i \rangle
       = \amgu\bigl(c_i, x = f(y, w)\bigr)
\),
where we have
$l'_1 = l'_2 = l'_3 = \VI$ and
$l'_4 = \{ w, z \}$.
In all four cases, we have $z \in l'_i$,
so that $z$ keeps its linearity even after
merging the results of the four subcomputations
into a single abstract description.

In contrast, when performing the same computation
with the original abstract description $\sfl$
in the decomposition phase, we also obtain a fifth component,
\begin{equation*}
  c_5 = \bigl\langle \{ w, xyz \}, f, l \bigr\rangle.
\end{equation*}
When computing
\(
  c'_5 = \langle \sh'_5, f'_5, l'_5 \rangle
       = \amgu\bigl(c_5, x = f(y, w)\bigr)
\),
we obtain $l'_5 = \{ w \}$, so that $z$ loses its linearity
when merging the five results into a single abstract description.
Note that this is not an avoidable precision loss,
since in the concrete computation path corresponding to
the substitution $\sigma$ we would have computed
\[
  \sigma'
    = \bigl\{
        x \mapsto f(x, w), 
        y \mapsto f(y, w), 
        z \mapsto f(z, w)
      \bigr\},
\]
where $z$ is bound to a non-linear term (namely, an infinite rational term
with an infinite number of occurrences of variable $w$).
Therefore, the result obtained when using the abstract description
$\sfl_\reduced$ is not correct.

As already observed in Section~\ref{sec:enhancedpos},
the above correctness problem lies not in the $\PSDFL$ domain itself,
but rather in our optimized implementation,
which removes the $\rho$-redundant elements
from the set-sharing description.

We implemented the first idea by Fil\'e
(i.e., the exploitation of property 1)
on the usual base domain $Pos \times \PSDFL$.
As noted above, this implementation may yield results
that are not correct: the precision comparison reported
in Table~\ref{tab:simplepos-vs-efree} provides an over-estimation
of the actual improvements that could be obtained
by a correct implementation.
However, it is not possible to assess the magnitude
of this over-estimation, since our implementation
of this enhancement on the domain $\Pos \times \SFL$,
where no $\rho$-redundancy elimination is performed,
times-out on a large fraction of the benchmarks.
The results in Table~\ref{tab:simplepos-vs-efree}
show that precision improvements are only observed
for goal-independent analysis.
When looking at the time comparisons, it should be observed that
the analysis of several programs had to be stopped
because of the combinatorial explosion in the decomposition,
even though we used the domain $\Pos \times \PSDFL$.
Among the proposals experimentally evaluated in this paper,
this one shows the worst trade-off between cost and precision.

Note that, in principle, such an approach to the recovery
of disjunctive information can be pursued beyond
the integration of sharing with freeness.
In fact, by exploiting the ground-or-free information
as in Section~\ref{sec:ground-or-free}, it is possible to
obtain decompositions where each component contains \emph{at most one}
occurrence (in contrast with the \emph{exactly one} occurrence of Fil\'e's
idea) of each ground-or-free variable.
In each component, the ground-or-free variable could then be
``promoted'' as either a ground variable (if it does not occur
in the sharing groups of that component) or as a free variable
(if it occurs in exactly one sharing group).

It would be interesting to experiment with
the second idea of Fil\'e.
However, such a goal would require a big implementation effort,
since at present there is no easy way to incorporate this enhancement
into the modular design of the \china{} analyzer.\footnote{%
Roughly speaking, the $\SFL$ component should be able to produce
some new (implicit) structural information and notify it
to the enclosing $\Pattern(\cdot)$ component, which would then
need to combine this information with the (explicit) structural information
already available. However, in order to be able to receive notifications
from its parameter, the $\Pattern(\cdot)$ component, which is implemented
as a \Cplusplus{} template, would have to be heavily modified.}

\section{Tracking Compoundness}
\label{sec:compoundness}

In \cite{BruynoogheCM94,BruynoogheCM94TR},
Bruynooghe and colleagues considered the combination of the standard
set-sharing, freeness, and linearity domains
with compoundness information.
As for freeness and linearity, compoundness was represented by
the set of variables that definitely have the corresponding property.

As discussed in \cite{BruynoogheCM94,BruynoogheCM94TR},
compoundness information is useful in its own right for clause indexing.
Here though, the focus is on improving sharing information,
so that the question to be answered is:
can the tracking of compoundness improve the sharing analysis itself?
This question is also considered in \cite{BruynoogheCM94,BruynoogheCM94TR}
where a technique is proposed that exploits the combination
of sharing, freeness and compoundness.
This technique relies on the presence of the occurs-check.

Informally, consider the binding $x=t$ together with
an abstract description where
$x$ is a free variable, $t$ is a compound term
and $x$ \emph{definitely} shares with $t$.
Since $x$ is free, $x$ is aliased to one of the variables occurring in $t$.
As a consequence, the execution of the binding $x=t$
will fail due to the occurs-check.
In a more general case, when only \emph{possible} sharing information
is available, the precision of the abstract description can be
safely improved by removing, just before computing the abstract binding,
all the sharing groups containing both $x$ and a variable in $t$.
In addition, if this reduction step removes all the sharing groups
containing a free variable, then it can be safely concluded that
the computation will fail.

To see how this works in practice,
consider the binding $x = f(y, z)$ and the description
$\sfl_1 \defeq \langle \sh_1, f_1, l_1 \rangle \in \SFL$ such that
\begin{align*}
  \sh_1 &\defeq \{ wx, xy, xz, y, z \}, \\
  f_1   &\defeq \{ x \}, \\
  l_1   &\defeq \{ w, x, y, z \}. \\
\intertext{%
Since $x$ is free and $f(y, z)$ is compound,
the sharing-groups $xy$ and $xz$ can be removed
so that the $\amgu$ computation will give
the set-sharing and linearity components
}
  \sh'_1 &\defeq \{ wxy, wxz \}, \\
  l'_1    &\defeq \{w, x, y, z \} \\
\intertext{%
  instead of the less precise
}
  \sh'_1 &\defeq \{ wxy, wxz, xy, xyz, xz \}, \\
  l'_1   &\defeq \{w\}.
\end{align*}
Note that the precision improvement of this particular example
could also be obtained by applying, in its full generality,
the second technique proposed by Fil\'e and sketched
in the previous section.
This is because the term with which $x$ is unified
is ``explicitly'' compound.
However, if the term $t$ was ``implicitly'' compound
(i.e., if it was an abstract variable known to represent compound terms)
then the technique by Fil\'e would not be applicable.
For example, consider the binding $x = y$
and the description
$\sfl_2 \defeq \langle \sh_2, f_2, l_2 \rangle \in \SFL$ such that
\begin{align*}
  \sh_2 &\defeq \{ wx, xyz, y \}, \\
  f_2   &\defeq \{ x \}, \\
  l_2   &\defeq \{ w, x, y, z \} \\
\intertext{%
supplemented by a compoundness component ensuring that
$y$ is compound.
Then the sharing-group $xyz$ can be removed
so that the $\amgu$ will compute
}
  \sh'_2 &\defeq \{ wxy \}, \\
  l'_2   &\defeq \{ w, x, y, z \} \\
\intertext{%
  instead of
}
  \sh'_2 &\defeq \{ wxy, wxyz, xyz \}, \\
  l'_2   &\defeq \{ w \}.
\end{align*}
To see how a knowledge of the compoundness
can be used to identify definite failure,
consider the unification $x = f(y, z)$ and the description
$\sfl_3 \defeq \langle \sh_3, f_3, l_3 \rangle \in \SFL$ such that
\begin{align*}
  \sh_3 &\defeq \{ wxy, wxz, x, y, z \}, \\
  f_3   &\defeq \{ w, x \}, \\
  l_3   &\defeq \{ w, x, y, z \}.
\end{align*}
As in the examples above, variable $x$ is free
and term $t \defeq f(y, z)$ is compound so that,
by applying the reduction step,
we can remove the sharing groups $wxy$ and $wxz$.
However, this has removed all the sharing groups containing
the free variable $w$, resulting in an inconsistent computation state.

We did not implement this technique, since it is only sound
for the analysis of systems performing the occurs-check,
whereas we are targeting at the analysis of systems
possibly omitting it.
Nonetheless, an experimental evaluation would be interesting
for assessing how much this precision improvement can affect
the accuracy of applications such as occurs-check reduction.

\section{Conclusion}
\label{sec:conclusion}

In this paper we have investigated eight enhanced
sharing analysis techniques that, at least in principle,
have the potential for improving the precision of the
sharing information over and above that obtainable
using the classical combination of set-sharing
with freeness and linearity information.
These techniques either make a better use of the already available
sharing information, by defining more powerful abstract semantic operators,
or combine this sharing information with that captured
by other domains.
Our work has been systematic since, to the best of our knowledge,
we have considered all the proposals that have appeared
in the literature:
that is, better exploitation of groundness, freeness,
linearity, compoundness, and structural information.

Using the \china{} analyzer,
seven of the eight enhancements have been experimentally evaluated.
Because of the availability of a very large benchmark suite,
including several programs of respectable size,
the precision results are as conclusive as possible
and provide an almost complete account of what is to be expected
when analyzing any real program using these domains.

The results demonstrate that good precision improvements
can be obtained with the inclusion of explicit structural information.
For the groundness domain $\Pos$, several good reasons have been given
as to why it should be combined with set-sharing.
As for the remaining proposals, it is hard to justify them
as far as the precision of the analysis is concerned.

Regarding the efficiency of the analysis,
it has been explained why the reported time comparisons
can be considered as upper bounds to the additional cost
required by the inclusion of each technique.
Moreover, it has been argued that, from this point of view,
the addition of a `ground-or-free' mode
and the more precise exploitation of linearity
are both interesting:
they are not likely to affect the cost of the analysis and,
when this is the case, they usually give rise to speed-ups.

No further positive indications can be derived from
the precision and time comparisons of the remaining techniques.
In particular, it has not been possible to identify a good heuristic
for the reordering of the non-grounding bindings.
The experimentation suggests that sensible precision
improvements cannot be expected from this technique.
When considering these negative results, the reader should be aware
that the precision gains are measured with respect to
an analysis tool built on the base domain
$\Pos \times \SFL$
which, to our knowledge, is the most accurate
sharing analysis tool ever implemented.

The experimentation reported in this paper resulted in
both positive and negative indications.
We believe that all of these will provide the right focus
in the design and development of useful tools for sharing analysis.

\section*{Acknowledgments}
This paper is dedicated to all those who take a visible stance
in favor of scientific integrity.
In particular, it is dedicated to David Goodstein, for
\emph{``Conduct and Misconduct in Science''};
to John Koza, for \emph{``A Peer Review of the Peer Reviewing Process
of the International Machine Learning Conference''};
to Krzsystof Apt, Veronica Dahl and Catuscia Palamidessi
for the Association for Logic Programming's
\emph{``Code of Conduct for Referees''};
and to the large number of honest and thorough referees
who do so much to help maintain and improve the quality of all
publications.


\hyphenation{ Ba-gna-ra Bie-li-ko-va Bruy-noo-ghe Common-Loops DeMich-iel
  Dober-kat Er-vier Fa-la-schi Fell-eisen Gam-ma Gem-Stone Glan-ville Gold-in
  Goos-sens Graph-Trace Grim-shaw Her-men-e-gil-do Hoeks-ma Hor-o-witz Kam-i-ko
  Kenn-e-dy Kess-ler Lisp-edit Lu-ba-chev-sky Nich-o-las Obern-dorf Ohsen-doth
  Par-log Para-sight Pega-Sys Pren-tice Pu-ru-sho-tha-man Ra-guid-eau Rich-ard
  Roe-ver Ros-en-krantz Ru-dolph SIG-OA SIG-PLAN SIG-SOFT SMALL-TALK Schee-vel
  Schlotz-hauer Schwartz-bach Sieg-fried Small-talk Spring-er Stroh-meier
  Thing-Lab Zhong-xiu Zac-ca-gni-ni Zaf-fa-nel-la Zo-lo }

\end{document}